# An overview of the structural correlation of magnetic and electrical properties of $Pr_2NiMnO_6$ Double perovskite

## ABSTRACT

Double perovskites $R_2NiMnO_6$ (R= Rare earth element) (RNMO) are a significant class of materials owing to their varied tunability of the magnetic and electrical properties with the structural modifications. $Pr_2NiMnO_6$ (PNMO) is one among the least explored members of this series, which shows spin phonon coupling, magnetocaloric effect and electrochemical performance for various applications such as spintronics, magnetocaloric refrigerant and solid oxide fuel cells. Most of the studies in PNMO is limited to the application domain and focus on the comparative study with different rare earth elements. Detailed structural studies like neutron diffraction are sparse in PNMO samples which will give a perception to the ordering in the compound that strongly depend on the physical and chemical properties. This review article goes through the various aspects of PNMO materials that have been reported till now and showcases the octahedral distortions and corresponding structural changes and the exchange interactions, which in turn correlates with the magnetic and electrical properties. The comparison study of PNMO with other members of the RNMO family and the relevance of PNMO over other members is also tried to showcase through this article. Hope this article provides an insight to the scope of studies in PNMO material for exploring unexposed properties of the materials in the double perovskite family.

## Introduction

Multiferroic materials with novel physical properties [1,2] are technologically useful and offer scopes in understanding rich physics. The perovskite oxide $ABO_3$ In which A site is 12 coordinated and the B site is octahedrally coordinated. The most interesting class of compounds with A position is occupied by rare earth or alkaline earth ions with large ionic radii and the B position is generally a transition metal (TM) ion or rare earth ion with smaller ionic radii that exhibit various physical properties. It contains corner sharing $BO_6$ octahedra and the voids are filled by the larger radius A cations. Fractional substitution at different extent at A and B lattice sites can make these compounds rich in variations in tuning the properties. In this the fraction of one type of cation is replaced by another cation. These substitutions are possible in A site alone, B site alone, and both A and B site cations. When these cations are ordered then the structure becomes double perovskite with formula $AA'B_2O_6$ [3] $A_2B'B''O_6$, $AA'B'B''O_6$ [4] . Two types of B cation ordering is possible mainly, that results in a double perovskite structure in a composition of $AB'_{1/2}B''_{1/2}O_3$ [5] or $A_2B'B''O_6$ as the substitution percentage at the B site reaches 50% [6,7] . The illustration of the $A_2B'B''O_6$ in the rock salt double perovskite structure in which the A cations are surrounded by an alternating network of $B'O_6$ and $B''O_6$ octahedra are depicted in the book, Perovskite: Crystallography, Chemistry and Catalytic Performance (Nova Publishers, 2013)[15] .Other kinds of double perovskites also occur, such as $AB'_{1/3}B''_{2/3}O_3$[8] $AB'_{1/4}B''_{3/4}O_3$[7]. The $BO_6$ octahedra can expand/contract and tilt for compensating the non-ideal ionic size ratios. Also, electronic instabilities may cause the cations to move from their ideal pseudo cubic positions or the distortions of octahedra can occur. In this double perovskite structure partial doping of other elements and creation of vacancies on any of the lattice sites can be possible. This flexibility enables most of the elements in the periodic table to occupy this perovskite lattice sites [9,10]. Because of the different electronic arrangements and ionic radii of TM ions, various properties can be studied by modulating the combinations and interactions among the TM ions. In $A_2B'B''O_6$ Perfect ordering of ions in B' and B'' sites is occasional and many physical and chemical properties are interrelated with the extent of ordering. A site can be occupied with $A^{1+}$(Alkali elements), $A^{2+}$ (Ca, Sr, Ba, Pb, Cd) and $A^{3+}$ (Rare earth elements) cations. $R_2B'B''O_6$ (R = rare earth, Sc, Y) have $R^{3+}$ cation and a combination of $B'^{1+}/B''^{5+}$, $B'^{2+}/B''^{4+}$, $B'^{3+}/B''^{3+}$, cations in the Double perovskite structure[7]. R cation is having coordination number 8. B cations are in octahedral coordination. Because of the large octahedral tilting some of the nearest neighbour anions get out of the first coordination sphere of the R cation. This is the reason behind the reduced coordination number of R is 8 rather than 12. An anion which is more than 3 Å far from A cation of the double perovskites can be considered to be outside the coordination sphere [11]. The research in area of A site ordering of this family of double perovskites is less compared to the B site ordering. The result of these rare studies of Asite ordering is particularly due to the fact that most of the materials with these sorts of ordering needs high pressure synthesis and exist over a very narrow range of temperature [12–14].

The studies of RNMO where A site is occupied with rare earth and the B site is with Ni and Mn ions is an interesting field of research. Recently the research related to each member of the RNMO family has flourished due to the extensive properties exhibited by the LNMO material. The magnetic and electrical properties in RNMO are structure correlated and in this review paper structural and physical properties and the various applications of double perovskite PNMO so far has been investigated.

## Parent compounds Single perovskites $PrMnO_3$ and $PrNiO_3$

In Rare-earth Nickelates $RNiO_3$ the band-structure shows a rich phase diagram, which can change through the parameters such as rare earth ionic radius, Ni–O–Ni bond angle, strain, etc [16,17]. Strongly correlated systems like rare earth manganites $RMnO_3$ also exhibit properties such as magnetocaloric effect, colossal magnetoresistance, charge-ordering and many other exciting properties as well [18–25]. Ni and Mn ions which are Jahn-Teller ions lead to the electronic and magnetic properties in these two perovskite systems.

The parent compounds of PNMO double perovskites are single perovskites $PrNiO_3$ and $PrMnO_3$ which are having Orthorhombic *Pbnm* structure with $Pr^{3+}$ occupying the A site and B sites are occupied by $Mn^{3+}$ in $PrMnO_3$ and $Ni^{3+}$ in $PrNiO_3$ respectively. $PrMnO_3$ is an A-type and $PrNiO_3$ is a G-type antiferromagnetic insulator, whereas $Pr_2NiMnO_6$ is a ferromagnetic insulator [26]. In $PrMnO_3$ the spins are ferromagnetically aligned in the basel plane and antiferromagnetic arrangement of spins in the perpendicular direction with Neel temperature at $T_N \approx 100$ K [27,28]. In $RNiO_3$ below the characteristic metal−insulator transitions the charge disproportionation of $Ni^{3+\delta}$ and $Ni^{3-\delta}$ is observed, [29] with δ being larger for the heavier lanthanides.[30]

Ni–O–Ni bond in $PrNiO_3$ and Mn–O–Mn bond in $PrMnO_3$ is having p band to d band interaction, which significantly influence physical properties of both compounds. Oxygen annealing is having a great impact on p band and d band overlap [31]. There is a sharp transition of metal (paramagnetic state) at high temperature to insulator (antiferromagnetic state) at lower temperature in $PrNiO_3$. This metal insulator transition temperature (TMI) is associated with significant change in conductivity [32]. TMI of $PrNiO_3$ is the same as the Neel temperature (TN). Hence these properties make $PrNiO_3$ and $PrMnO_3$ as perfect materials for exploring the structure correlated magnetic and electrical properties [31]. Many studies have been reported for single perovskite $PrMnO_3$ [27,33–40] and $PrNiO_3$ [32,41–44], A site modified $PrMnO_3$ [18–25,45–49] and $PrNiO_3$ [50,51], B site modified $PrMnO_3$ [52,53] and $PrNiO_3$ [32,54] and A and B site modified $PrMnO_3$[55–61] and $PrNiO_3$ [62], theoretical studies on $PrMnO_3$ [38–40] and $PrNiO_3$ [43], structural and optical studies [27,34–36,41,57] for various applications such as catalysis [37,40,55] solid oxide fuel cells, [33,46] Colossal Magnetoresistance, [23,48] Colossal Dielectric Response [47], Multiferroicity [24,42], Ferroelectricity [19], Ferromagnetism [18,20], Antiferromagnetism [56], Magneto caloric effect [25,58,60], Unusual metamagnetism [59], Thermoelectric property [61], Water oxidation [44] etc.

**Combination of different elements at B'/B" site with Pr at A site in double perovskite $R_2$ B' B"$O_6$**

The double perovskite structure can be influenced with the B-site cation ordering, bond covalency, ionic radii of cations, electronic instabilities, bonding preferences, oxygen content and synthesis conditions [7]. As a general rule a disordered arrangement is possible when the oxidation state difference of B' and B" is less than two (E.g., $Pr_2CrFeO_6$), whereas, if it differ by greater than two, nearly always leads to an ordered arrangement (E.g., $Pr_2LiRuO_6$). When the oxidation state difference is two (E.g., $Pr_2NiMnO_6$), disordered, partially ordered, or fully ordered arrangements occur based on differences in size and/or preference of bonding of the B' and B" ions [11].

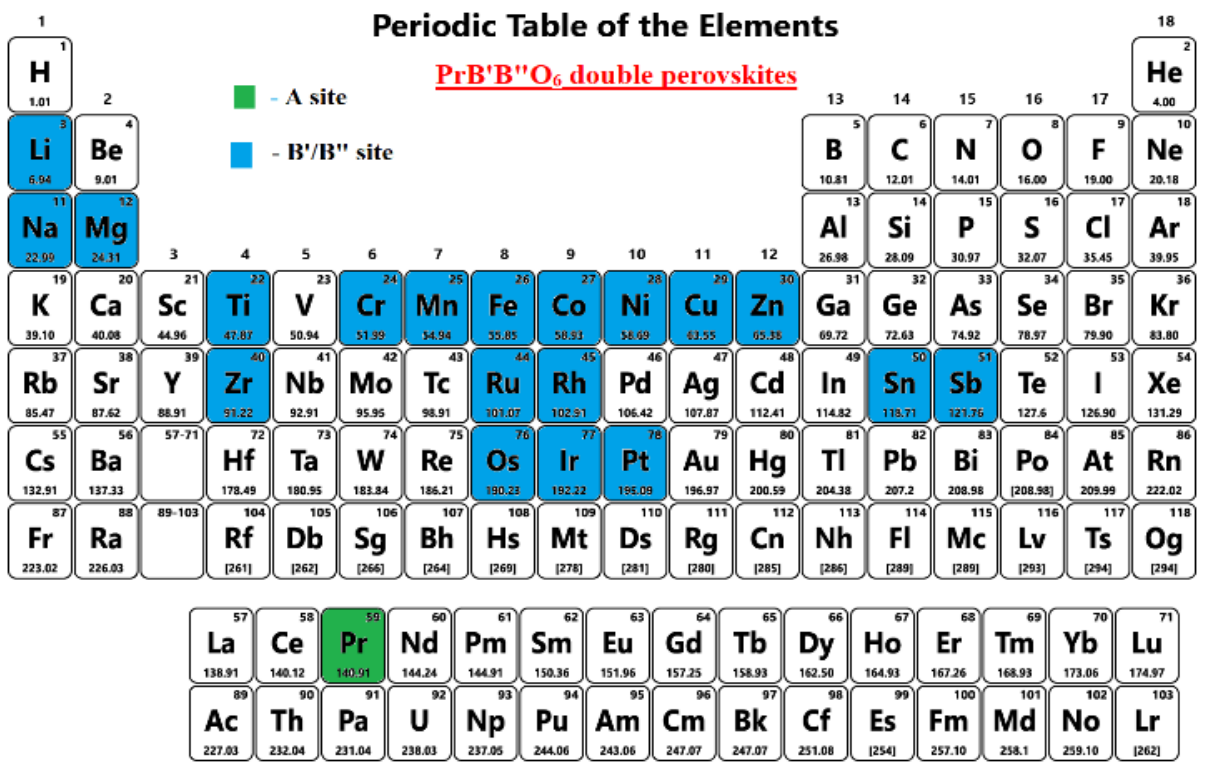

*Figure1: The choice of different elements in the B' and B" sites with Pr at A site in the double perovskite structure.*

Double perovskite structure with Pr at A site and different combinations of elements at B' and B" sites are studied with various groups and those available in literature is tabulated in Table1. Different elements can occupy the B site in different combinations to form the $Pr_2B'B"O_6$ double perovskite structure (Fig:2). The periodic arrangement of these ions gives significant effects to the conductivity, thermal expansion behavior and catalytic properties apart from changes in the magnetic and electrical properties [63–65].

Depending on the element in the B'/ B" site there are changes in the exchange interaction of these compounds. The charge difference and the size difference are having a significant role in the arrangement of the B cation sublattice. As the charge and size difference among B cations increases, the stability of rock salt arrangement becomes more stable than random arrangement. Anderson *et al.* in 1993 reported that the main criteria for ordering in perovskite compounds are the charge difference, size, electronic configuration of B-cation and the size mismatch between A/B cations [66]. Depending on the

| Summary of all known double perovskite with Pr at A site and different combination of elements at B' and B" sites | | |
|---|---|---|
| **A site** | **B' site** | **B" Site** |
| Pr | Li | Ir [67], Ru [68], Os [69] |
| | Na | Ir [70], Pt , Sn, Ti [71], Os [72], Ru [73] |
| | Mn | Rh[74], Ti [75], Co [76–79], Cr [80,81], Sb [82] |
| | Cu | Sn, Zr [83], Ir [66], Ti [84] |
| | Ni | Pt [85], Ru [71], Ir [86], Mn [26,87,88], Zr [89] |
| | Co | Pt [85], Ru [90] |
| | Zn | Pt [85], Ti [91], Zr [92] |
| | Fe | Cr [93,94] |
| | Mg | Ir [67,86], Pt[85] |

superexchange interaction between the B' and B" cations the material shows ferromagnetic or antiferromagnetic ordering. Also, the presence of different magnetic ions in the lattice sites create various magnetic interactions which leads to the canted spins in the lattice. These canted spins create additional magnetic transitions observed in the thermo magnetization curves. The structural variations and the differences in the exchange interactions of different ions at B' and B" sites with Pr at A site is tabulated in Table 2.

## The relevance of Ni and Mn at B' and B" site

RNMO materials are a distinct class of double perovskite oxides, in which $e_g^2$ -O- $e_g^0$ electronic interaction leads d electron spins to align ferromagnetically based on the classical Goodenough–Kanamori rules [95,96]. In $La_2NiMnO_6$(LNMO) a ferromagnetic transition temperature ($T_C$ = 280 K) which is near room temperature attracted considerable attention to this class of compounds [10,97]. Numerous properties have been explored lately in this system such as magnetocapacitance [10,98] colossal magnetoresistance [99–101], super conductivity [102,103], magnetodielectric properties [10], catalytic property [104–106], multiferroicity [1,2,107,108]. Novel technologically advanced devices like spin filter junctions and multiple state memory elements can be designed with the utilization of the complex physical properties of RNMO materials [109–113]. Also, they provide different opportunities to induce multiferroicity and its modulation in oxide systems.

| Table 2: structural, exchange interaction and curie temperature of double perovskite with Pr AT a site and different combination of TM at B'/B" site | | | | | | | |
|---|---|---|---|---|---|---|---|
| Material | Space group | oxidation state | Bond length | Bond angle | $\beta$ | Curie temperature (Tc)/Neel temperature (TN) | Superexchange interaction |
| $Pr_2MgIrO_6$ | $P2_1/n$ [$a^- a^- b^+$] | $Mg^{2+}$, $Ir^{4+}$ | Mg-O =2.058 Ir-O =2.014 | 150.27 | 90.006 | $T_N$ = 14 K | Antiferromagnetic interaction |
| $Pr_2NaRuO_6$ | $P2_1/n$ [$a^- a^- b^+$] | $Na^+$, $Ru^{5+}$ | Na-O=2.278 Ru-O=1.954 | 142.93 | 90.786 | $T_N$ = ~20 K | Antiferromagnetic interaction (spin canted weak ferromagnetism) |
| $Pr_2LiIrO_6$ | $P2_1/n$ [$a^- a^- b^+$] | $Li^+$, $Ir^{5+}$ | Li-O=2.106 Ir-O = 1.970 | 149.43 | 90.222 | $T_N$ = ~57 K | Antiferromagnetic interaction |
| $Pr_2LiOsO_6$ | $P2_1/n$ [$a^- a^- b^+$] | $Li^+$, $Os^{5+}$ | Li-O = 2.113 Os-O = 1.963 | 148.9 | 90.407 | $T_N$ =35 K | Antiferromagnetic interaction |
| $Pr_2LiRuO_6$ | $P2_1/n$ [$a^- a^- b^+$] | $Li^+$, $Ru^{5+}$ | Li-O= 2.103 Ru-O= 1.950 | 150 | 90.285 | $T_N$ =31 K | Antiferromagnetic interaction |
| $Pr_2NiIrO_6$ | $P2_1/n$ [$a^- a^- b^+$] | $Ni^{2+}$, $Ir^{4+}$ | Ni-O = 2.063 Ir-O = 2.012 | 149.4 | 90.005 | $T_N$ = 105k | Antiferromagnetic interaction (spin canted weak ferromagnetism) |
| $Pr_2NaIrO_6$ | $P2_1/n$ | $Na^+$, $Ir^{5+}$ | Na-O = 2.282 Ir-O = 1.972 | 142.37 | 90.793 | - | absent |
| $Pr_2CuSnO_6$ | $P2_1/m$ | $Cu^{2+}$, $Sn^{4+}$ | Cu-O = 2.086 | 156 | 91.716 | $T_{N1}$ =220 K $T_{N2}$ = below 40 K | Antiferromagnetic interaction (spin canted weak ferromagnetism) |
| $Pr_2FeCrO_6$ | R3c | $Fe^{3+}$, $Cr^{3+}$ | - | - | - | $T_C$= 562 K | Ferromagnetism |
| $Pr_2FeCrO_6$ | Pbnm [$a^+a^-b^-$] | $Fe^{3+}$, $Cr^{3+}$ | Fe/Cr -O= 1.753 | 159.77 | - | $T_C$= 240 K | Ferromagnetism (spin canted weak antiferromagnetism) |
| $Pr_2CrMnO_6$ | pbnm | $Cr^{2+}$, $Mn^{4+}$ | - | 159.73 | 90 | $T_C$ = 93.74 | Ferromagnetic interaction |
| $Pr_2CoMnO_6$ | $P2_1/n$ | $Co^{2+}$, $Mn^{4+}$ | - | - | - | $T_C$ =173 K | Ferromagnetic interaction |
| $Pr_2NaOsO_6$ | $P2_1/n$ [$a^- a^- b^+$] | $Na^+$, $Os^{5+}$ | Na-O = 2.291 Os-O = 1.967 | - | 90.854 | $T_{N1}$=20 K $T_{N2}$ =10 K | Antiferromagnetic interaction |
| $Pr_2CoRuO_6$ | $P2_1/n$ | $Co^{2+}$, $Ru^{4+}$ | Ru/Co-O= 2.046 | 146.53 | 90.005 | No magnetic studies | |
| $PrMn_{1/2}Rh_{1/2}O_3$ | pbnm | $Mn^{3+}$ $Rh^{3+}$ | Mn/Rh–O=2.03 | 150.33 | | No magnetic studies | |
| $Pr_2ZnTiO_6$ | $P2_1/n$ [$a^+b^-b^-$] | $Zn^{2+}$, $Ti^{4+}$ | Zn-O=2.049 Ti-O=1.955 | 152.71 | 90.264 | No magnetic studies | |
| $Pr_2NiZrO_6$ | $P2_1/n$ | $Ni^{2+}$, $Zr^{4+}$ | Ni-O=1.926 Zr-O=2.267 | 145 | 90.059 | No magnetic studies | |

In the double perovskite $A_2B'B''O_6$ shows a wealth of structural variations due to the presence of three cations to influence the magnetism. There are various possibilities for the occupancy of magnetic cation only at B' site (nonmagnetic ions at B'' and A site), only at A site, only at B'' site, both B' and B'' sites, and finally, the A, B' and B'' sites. For choice of A site cation large ionic radii with magnetic property is needed. There is a great selection of trivalent cations in trivalent lanthanides as ideal candidates to take the role of A site cations. After the selection of lanthanides (Ln) for A site then the compound is $Ln_2B'B''O_6$. In this compound the sum of the B' and B'' oxidation states are limited to +6 (+1/+5, + 2/+4, and +3/+3). When the number of magnetic ions increases the magnetic behaviour of the double perovskite structure becomes complex [86]. In the lanthanide series $La^{3+}$ is a nonmagnetic ion, $Pr^{3+}$ is the largest magnetic ion in the lanthanide series after $Ce^{3+}$. Moreover, from the viewpoint of magnetism, when the nonmagnetic La ion is replaced with the magnetic Pr ion the various magnetic interactions arise will lead to another dimension of the application of these compounds with richer phase diagrams.

Once the settlement of $Pr^{3+}$ in the A site, next is the choice of elements to B'/B'' sites. The B site ordering is mainly influenced by the charge difference of B' and B'' cations ($\Delta Z_B$) and the size difference of these ions ($\Delta r_B$). For $\Delta Z_B > 2$ leads to highly ordered double perovskite structure, whereas $\Delta Z_B < 2$ leads to disordered samples[7]. In the case of $\Delta r_B$, the larger the size difference most probable is the ordering. Most interesting class of compounds are $\Delta Z_B = 2$, where order-disorder phenomena are observed. Here a wide range of partial ordering is taking place. Also, for a wide range of size differences both ordered and disordered states are observed, but for small $\Delta r_B$ disordered states are most favourable. The ordering of B'/ B'' octahedra will be favourable if the size difference is greater than 0.1 Å and charge difference is two or greater than two [91]. RNMO family exhibit highly ordered structure with magnetic TM cations $Ni^{2+}$ and $Mn^{4+}$ having a charge difference of 2 and size difference $\Delta r_B = 0.13\text{-}0.16$ Å. Also, the ferromagnetic superexchange interaction between these cations also increases the criteria of selecting $Ni^{2+}$ and $Mn^{4+}$ to the B'/B'' sites. For compounds having $\Delta Z_B = 2$ other factors which influence the ordering are synthesis conditions. The annealing temperature and the synthesise pressure will influence the ordering. Another weak effect is the smaller the A site cation larger the B site ordering. This is because smaller A site cations bring the B' and B'' cations closer, which increases the repulsion between them. That leads to the increase in Madelung energy of ordering. Finally, oxygen vacancies also influence the ordering. The compounds with oxygen vacancies are of lesser B site cation order than oxygen stoichiometric compounds [114]. In RNMO the ordering of B sites can be done with $Ni^{3}/Mn^{3+}$ or $Ni^{2+}/Mn^{4+}$. The rock-salt ordering between $Ni^{3+}$ and $Mn^{3+}$ would be derived from the Jahn−Teller character of $Mn^{3+}$, while $Ni^{2+}/Mn^{4+}$ order would be due to charge ordering. The studies with NMR and XAS [115–117] show the actual valence states are $Ni^{2+}$ and $Mn^{4+}$. The charge ordering of these ions leads to the attractive physical properties of these double perovskites.

Apart from the extensive studied LNMO, studies of other rare earth RNMO oxides are comparatively scarce [118–122]. Extensive research is carried out for decades for LNMO in the areas of A site doped [123–125], B site doped [126–128], A and B site doped [129] and theoretical studies [130,131] for various technological applications [132]. Because of these multiple properties Researchers are curious about the modulation of these exciting properties with the change of rare earth element. Hence the studies are extended to other members of the RNMO family as well. Even though there are many similarities and regular trends in the physical properties shown by the members of RNMO family there are many differences in the properties which are related with the particular rare earth ion and the coupling between rare earth ion and the Ni--Mn system. These differences can be realized only by exploring each member of the RNMO family through extensive research and correlation studies. The individual studies on different members of the RNMO family also started to explore the physical properties of this distinct group of compounds. Apart from LNMO and PNMO other rare earth ( Lanthanides (La to Lu) and Scandium, Yttrium) double perovskites $Ce_2NiMnO_6$ [133], $Nd_2NiMnO_6$ (NNMO) [134,135], $Pm_2NiMnO_6$, $Sm_2NiMnO_6$ (SNMO) [136–138], $Eu_2NiMnO_6$ (ENMO) [139–141] , $Gd_2NiMnO_6$ (GNMO) [142–144], $Tb_2NiMnO_6$ (TNMO) [145,146], $Dy_2NiMnO_6$ (DNMO) [147–151], $Ho_2NiMnO_6$ (HNMO) [152,153], $Er_2NiMnO_6$ (ENMO) [153,154], $Tm_2NiMnO_6$ (TNMO) [155], $Yb_2NiMnO_6$ (YNMO) [108], $Lu_2NiMnO_6$ (LNMO) [156–159], $Sc_2NiMnO_6$ [160], $Y_2NiMnO_6$ [161–163] are also now on the path of research.

Due to the comparable size of $Mn^{4+}$ and $Ni^{2+}$, the charge difference of 2, which helps to tune ordering of B'/B" site cations, covalent nature of bonds and superexchange interactions in the compound. TM ions Ni and Mn exhibit a paramount role for influencing the various characteristics exhibited by RNMO double perovskites. Hence Ni/Mn combination at B'/B"-site in $R_2B'B''O_6$ double perovskite structure brings exciting magnetic and electrical interactions resulting in the formation of new phase or lead to entirely different behaviours. In single perovskite $PrNiO_3$ and $PrMnO_3$ the interactions Ni-Ni and Mn-Mn are antiferromagnetic in nature, while oxide systems having both Mn and Ni ions like RNMO there may be the possibility of competition between different magnetic phases [31].

In PNMO $Ni^{2+}$ and $Mn^{4+}$ occupy the B' and B" site in an alternate octahedral site, where the charge difference is 2. Blasse *et al.* [164] explored the magnetic properties of perovskite $La_2MeMnO_6$ (Me = Mg, Co, Ni, Cu), where Me has +2 oxidation state and Mn is in +4 state. The magnetic interactions between paramagnetic ions can be ferromagnetic or antiferromagnetic in non-metallic oxides. Goodenough [96] and Kanamori [95] rules have introduced a set of semiempirical rules to realise these magnetic interactions. The sign of these interactions had been determined earlier by Jonker [165]. The $180^0$ (the lines connecting the interacting cations to the intervening anion make an angle $180^0$) $Me^{2+}$- O - $Mn^{4+}$ magnetic interaction through oxygen ions seem to be positive. PNMO double perovskite structured material reported magneto-dielectric, magnetocaloric, spin phonon coupling and electrochemical performance so far, for applications in the field of spintronics [88,166,167], magnetic refrigeration [88,152], Solid oxide fuel cells [168,169] etc.

**Superexchange interaction in RNMO**

Super exchange interaction is derived from the second order perturbation theory. This interaction happens between two magnetic ions mediated through a nonmagnetic ion such as oxygen. This magnetic interaction strongly depends on the Metal – oxygen - metal bond angle. The magnetic interaction through super exchange was introduced by Goodenough [96][12]. Tiana *et al.*[170] pictured in figure:2 about how these super exchange interactions induce ferromagnetism or anti-ferromagnetism in a material. This ferro/anti-ferromagnetism is governed by the Pauli's exclusion principle. Hence in PNMO $Mn^{4+}$ and $Ni^{2+}$ magnetic cations interact via nonmagnetic $O^{2-}$ anions through Mn-O-Ni bond. Since the tilting of $MnO_6$ and $NiO_6$ octahedra will reduce the Mn-O-Ni bond angle which will affect this superexchange interaction. $Mn^{4+}$ and $Ni^{2+}$ is in $3d^8$ and $3d^3$ configurations. The oxygen octahedral environment around Ni and Mn ions splits the d levels in $t_{2g}$ and $e_g$ states. The crystal field splitting and the spin splitting are to different extent for Ni and Mn ions. For Mn site the crystal field splitting is smaller, while at Ni site crystal field splitting is larger in comparison with the spin splitting in both ions. This leaves Ni $t_{2g}$ states to be completely filled, Ni $e_g$ states and Mn $t_{2g}$ states being half-filled with Mn $e_g$ states being empty. Super exchange processes between Ni-$e_g$ state and Mn-$t_{2g}$ exhibit anti ferromagnetism. However, Ni-$e_g$ and Mn-$e_g$ states show ferromagnetism based on Hund's rule coupling $J_H$. Das *et al.*[171] in figure:6 illustrate AFM and FM super exchange paths between Ni-$e_g$ spin (S = 1) and Mn-$t_{2g}$ spin (S = 3/2) and Ni-$e_g$ spin (S = 1) and empty Mn-$e_g$. The net magnetic interaction $J_{Ni–Mn}$ is expressed by considering the contributions, in terms of hopping interaction between Ni-$e_g$ and Mn-$e_g$ ($t_{e–e}$), Ni-$e_g$ and Mn-$t_{2g}$ hopping interaction ($t_{t–e}$), separation in energy levels of Ni-$e_g$ – Mn-$e_g$ ($\Delta_{e–e}$) and Ni-$e_g$ –Mn-$t_{2g}$ ($\Delta_{t–e}$), the choice of U and $J_H$ values. The tilting of the $NiO_6$ and $MnO_6$ octahedra makes $t_{t–e}$ non-zero, resulting in the variation of Ni–O–Mn angle, from 180°. Electron hopping in $e_g$ - $e_g$ orbital is larger than $e_g$ - $t_{2g}$. Also, the states of Ni-$e_g$ and Mn-$t_{2g}$ lie closer in energy than Ni-$e_g$ Mn-$e_g$ .[171].

Previous theoretical studies on LNMO demonstrated [171–173] both $t_{2g}$ and $e_g$ states in the majority spin channel were filled for Ni ions in the $d^8$ state (S = 1) spin configuration, and only the $t_{2g}$ states were occupied in the minority spin channel. Whereas, for the $Mn^{4+}$ in $d^3$ configuration (S = 3/2), only the $t_{2g}$ states were fully occupied in the majority spin channel, and both the $t_{2g}$ and $e_g$ states remained empty in the minority spin channel. Only the states with same symmetry and spin are allowed to interact in the superexchange interaction. Interactions between the Ni $e_g$ and Mn $e_g$ symmetries, (filled Ni eg states and the empty Mn $e_g$ states in majority spin channel) and those between the Ni $t_{2g}$ and Mn $t_{2g}$ symmetries (filled Ni $t_{2g}$ states and the empty Mn $t_{2g}$ states in minority spin channel) leads to ferromagnetic interaction according to Goodenough–Kanamori rules.

**Comparison of PNMO with other members of the RNMO family (R = La to Lu (Lanthanides), Ce, Sc, Y) family**

In $R_2B'B''O_6$ three cation orderings are possible. (1) The random arrangement of B' and B'' cations lead to $R_2B'_{1/2}B''_{½}O_3$. This disordered perovskite sample displays an orthorhombic symmetry with the space group of *Pbnm* at room temperature; (2) Ordered arrangement of B' and B'' cations resulting in a double perovskite $R_2B'B''$ $O_6$. This sample crystallizes in monoclinic symmetry (P$2_1/n$) at room temperature; (3) Many systems, mixed cation ordering exists [174] . The B'/B'' cation ordering extends to a few unit cells or domains. Also, all of these three phases are thermodynamically stable [175].

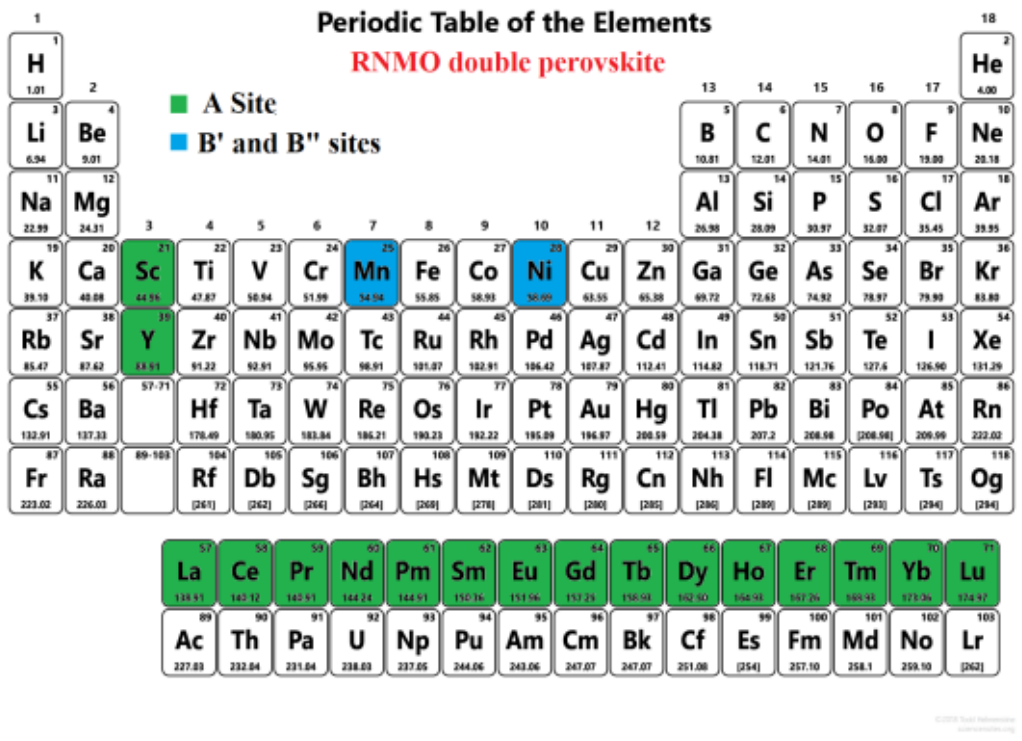

*Figure 2: The choice of different rare earth element in the A site of RNMO double*

In the A site of RNMO the trivalent rare earth ions Lanthanides (La to Lu) and Sc and Y can occupy (Fig:5). Decrement in the rare earth ionic radius ($r_R^{3+}$) in RNMO leads to the decrease in the unit cell volume. The lattice parameters a and c decrease while the b parameter more or less remains constant [121,122,176] with decrease in $r_R^{3+}$. Also, the value $c/\sqrt{2}$ lies between a and b. The primary force of distortions in this perovskite steric arises due to the reduction in the size of $R^{3+}$ cation.[155] The R-O distances decrease as the size of rare earth decreases. Mn/Ni–O bond lengths increase as the value of $r_R^{3+}$ decreases [121,122]. Whereas the average Ni/Mn-O bond lengths do not have significant variation with the $r_R^{3+}$ [122,155]. While the Ni-O-Mn bond angle decreases as $R^{3+}$ radius decreases (Table3). In a study of the comparison with the different RNMO compounds a difference concerning ⟨Ni−O⟩ and ⟨Mn−O⟩ bond lengths have observed. Smaller R cation occupancy to the A site introduced ⟨Ni−O⟩ distances are substantially higher whereas ⟨Mn−O⟩ distances are smaller.[155] This indicates the charge difference between Ni and Mn is large. Also, this would seem that the charge disproportionation between $Ni^{2+}$ and $Mn^{4+}$ stabilized more effectively with the rare earths having smaller ionic radii.[155] The tilting of the octahedra along the c axis results in the displaced oxygen atom to occupy in between the octahedra. Though octahedral distortions are small, there is a significant variation in the tilting angle of octahedra as $r_R^{3+}$ decreases.

Stability and distortion of perovskite crystal structure can be expressed by Goldschmidt tolerance factor (t) as;

$$t = \frac{r_R + r_O}{\sqrt{2}\left[\frac{r_{Ni} + r_{Mn}}{2} + r_O\right]}$$

$r_R$ , $r_O$, $r_{Ni}$, and $r_{Mn}$ indicates the effective ionic radii of rare earth, oxygen, Ni, and Mn ions correspondingly [177]. For $t < 1$ the A cation radius is less than ideal A site cation and the tilting of $BO_6$ octahedra and/or the change of bond length compensate for the size mismatch of this double perovskite structure. For t >1 A site cation is very large and the size mismatch cannot be overcome by the octahedral tilting. For that type of compounds various hexagonal structures are formed rather than perovskite structure[7] . For the trivalent rare earth ions, the ionic radii is smaller than the ideal value of the A site cations which leads to condition t < 1 and exhibit octahedral tilting. The value of tolerance factor decreases steadily with decrement in rare earth ionic radii which is interrelated with the electronic and magnetic properties [120,121].

The distinct decrement of Rare earth ion radius leads to the reduction in the Ni–O–Mn angle, which determines various characteristics of $RMO_3$ perovskite oxides [178] as well as RNMO double perovskites [179]. In RNMO as the rare earth ionic radius decreases the magnetic and electrical properties changes due to the alteration of superexchange interaction, which is closely related with Ni-O-Mn bond angle. There is a subsequent decrease in Tc as the value of $r_R^{3+}$decreases from La to Lu. Asai *et al.* [179] in a study of $REMe_{0.5}Mn_{0.5}O_3$ (RE = Rare earth element, Me = Co. Ni) found that $T_C$ of this system completely determined by superexchange interaction between $Mn^{4+}$ and $Me^{2+}$ as

$$T_c = \frac{2zJ_{Me-Mn}\sqrt{S_{Me}(S_{Me}+1)S_{Mn}(S_{Mn}+1)}}{3K_B}$$

Where J is the super exchange integral, S represents the spins of ions and Z is the distance between $Me^{2+}$ and $O^{2-}$ ions.

From this relation it is clear that the decrement of $T_c$ in this type of perovskite system is not because of the magnetic interaction of rare earth ions. A geometrical factor, super exchange Ni - O – Mn bond angle is playing a major role in the value of $T_C$. This superexchange angle depends on the kind of rare earth element in the double perovskite. Hence the decrease of the critical temperature of PNMO compared to LNMO is due to the decrease in the super exchange <Ni - O – Mn> angle from $180^0$ [121,176].

The variation of Tc is related with the tilting angle (ϕ) of octahedron. There is linear relationship between Tc and $\cos^2\phi$. Where ϕ is determined from the bond angle as 180 - <Ni–O–Mn>/2 [179]. There is a better correlation between $T_C$ and $R^{3+}$ radius [180] than with ϕ alone. From $^{55}Mn$ NMR experiments, Asai et al. [181] have shown that for the RNMO oxides a decrease of covalency of Ni–O and Mn–O bonds is observed when R goes from La to Lu. $T_C$ is controlled not only by the angle ϕ but also by the variation of covalency/ionicity of the Ni–O and Mn–O bonds. $R^{3+}$ radius seems to reflect both these effects and hence provides a better correlation for the $T_C$ [122]. In addition to this tilting of octahedra, the decrease of covalency of Ni/Mn-O bonds is also having a significant role in the reduction in the transition temperature as the size of the rare earth decreases. This is explained in one of the earliest studies using NMR hyperfine magnetic field [181].

There is a canted spin arrangement of $Ni^{2+}$ and $Mn^{4+}$ within the ferromagnetic arrangement of RNMO with R ions having lower ionic radii (R= Tb, Ho, Er, Tm). Which gives rise to a net ferrimagnetic arrangement [155]. By Goodenough− Kanamori rules the $180^0$ superexchange interaction $Ni^{2+}$-O-$Mn^{4+}$ path the sign of the interaction of Ni $e_g$ and Mn $e_g$ ferromagnetic interaction changes to antiferromagnetic as the angle becomes closer to 90° [95]. The angle decreases as the rare earth ion radii decreases in RNMO. Along with the decrease in angle the evolution of interaction from ferro to antiferromagnetic state makes an intermediate state for these compounds. Also, the magnetic superexchange interactions between perfectly parallel or antiparallel spins are stronger than between spins with different directions. Hence the presence of a major ferromagnetic phase with magnetic frustrations within it leads to a glassy state.

The various physical properties of RNMO materials, like the band gap, saturation magnetization, Curie point ($T_C$), changes noticeably in different reports [121,122,182]. The probable mechanism that leads to magnetism in these materials drew diverse conclusions. various synthesis routes that affect the level of the antisite defect at the Ni/Mn sites and the crystallographic structures may be one of the reasons for these differences in properties. variations in the local atomic structure about the Ni and Mn ions is also having an important role in determining the various physical properties [183,184].

Modifications on the crystal structure that induce changes in magnetic interactions can be done through replacing the A site with different rare earth ions. There are reports like magnetic order of RNMO samples would change from ferromagnetic to E*-type antiferromagnetic for R = La, Sm to R=Y respectively using DFT calculations. The E*-type magnetic structure (↑↑↓↓) allows a ferroelectric polarization to happen, resulting in multiferroicity [130]. However, the magnetic measurements show a definite ferromagnetic character of YNMO with specific $T_C$. It is reported that E*-type (↑↑↓↓) spin arrangements in YNMO exist only at zero magnetic field and this AFM arrangement can be destroyed with application of magnetic field.[185] This is the reason behind the proposed AFM arrangement is not exhibited by YNMO samples. [162] The existence of antiferromagnetic interactions on $Tb_2NiMnO_6$, $Nd_2NiMnO_6$ , $Sm_2NiMnO_6$ , $Pr_2NiMnO_6$ is possible in the studies carried out in these material [186,187]. As expressed by figure:1 by Zhao et al.[121] the experimentally determined correlation of the rare earth ionic radius with (a) $\beta$ and tolerance factor (b)Ni-O-Mn angle (c) Bandgap and curie temperature(d) Ni/Mn- Bond length are helpful for understanding the correlation of structure and physical properties.

According to the comparison table of RNMO (table 3) There is a presence of secondary magnetic transitions in some samples. In some of the reports these secondary transitions are absent in the same samples. This second transition at lower temperature arises due to the AFM interaction of the antisite defects. Ni-O-Ni and Mn-O-MN interactions are antiferromagnetic in nature. These weak AFM interactions will lead to the secondary transition at lower temperature. Also, there is a magnetic anomaly observed at very low temperature in the magnetisation curves. For Nd, Sm there is a downturn in FC magnetisation, while for Gd, Tb, Dy Ho an upturn is observed [122,188]. There are reports which claim low temperature anomalies of RNMO samples [121,177]. There are various explanations for this magnetic anomaly and these are still in debate. There are reports which claims this anomaly is due to the antiferromagnetic coupling of the Nd, Sm spins to the Ni/Mn sublattice and the ferromagnetic coupling between Gd, Tb, Dy spins with Ni/Mn sublattice. The magnetic moment of these rare earths is significant only at lower temperature and the moment will follow Hund's rule depending on the less than half filled (Nd, Sm) or more than half filled (Gd, Tb, Dy) conditions. The spin orbit coupling of rare earth and the Ni/Mn sublattice will lead to this low temperature variation in magnetisation [188]. Another claim for this anomaly is that there is no role for the R and Ni/Mn sublattice in these effects. This downturn is observed for RNMO systems irrespective of whether the R ion is magnetic or not. Because this anomaly is observed for YNMO samples where $Y^{3+}$ is a nonmagnetic ion. In such systems this low temperature magnetic anomaly is explained in terms of the random FM and AFM interactions that arise due to the variable oxidation state of Ni and Mn. The random AFM interactions $Ni^{3+}/Mn^{3+}$ - O - $Ni^{3+}/Mn^{3+}$ and FM interaction $Ni^{2+}$- O – $Mn^{4+}$ along with crystallographic disorder leads to this low temperature anomaly and the glassy state of these oxides [177] . Another possibility arises due to the antiphase AFM domains resulting from spatial distribution of $Ni^{2+}/Mn^{4+}$ ordered domains and the out-of-phase ordering [189].

**Table 3 : Structural and magnetic properties of RNMO (Lanthanides,Y) with space group $P2_1/n$**

| Double perovskite | Ionic radii (Å) | Space group | Bond length (Å) | Bond angle ($^0$) | $\beta$ ($^0$) | Magnetic transition temperature (K) | Exchange interaction |
|---|---|---|---|---|---|---|---|
| $La_2NiMnO_6$ 10,121,190 | 1.16 | $P2_1/n$ | Ni-O = 1.97 – 2.02<br>Mn-O = 1.917- 1.96 | 160.7-164.3 | 89.82-90.01 | $T_C$ - 273,240,280, 266<br>$T_C'$- 160,138,102 | Ferromagnetic interaction ($T_C$)<br>Antiferromagnetic interaction at antisite defects/antiphase boundaries ($T_C'$)<br>Anomaly in magnetisation curve ( Downturn in magnetisation for Nd, Sm and Upturn in magnetisation Gd, Tb, Dy ) ($T_2$) |
| $Ce_2NiMnO_6$[133] | 1.143 | | - | - | - | - | |
| $Nd_2NiMnO_6$ 121,122,135,152,174,187,191 | 1.11 | $P2_1/n$ | Ni-O = 2.04<br>Mn-O = 1.915 | 155-158 | 90.02 - 90.05 | $T_C$ =191- 200<br>$T_C'$ = ~ 100<br>$T_2$ = < 50 | |
| $Pm_2NiMnO_6$ | 1.09 | | All of its isotopes are radioactive | | | | |
| $Sm_2NiMnO_6$ 136–138,167,167,182,186 | 1.08 | $P2_1/n$ | Ni-O = 2.01-2.08<br>Mn-O = 1.88-1.93 | 151-154 | 89.8 - 90.06 | $T_C$-141 – 165<br>$T_C'$- 67<br>$T_2$ = < 20 | |
| $Eu_2NiMnO_6$ 140,141,177 | 1.066 | $P2_1/n$ | Ni-O = 1.88-2.01<br>Mn-O = 1.99-1.96 | 148.4 - 151.32 | 90.01-90.07 | $T_C$-145<br>$T_2$ - 6 | |
| $Gd_2NiMnO_6$ 122,177,188,192,193 | 1.05 | $P2_1/n$ | Ni-O = 2.070-2.19<br>Mn-O = 1.895-2.03 | 148.6 - 152.20 | 90.13 - 90.19 | $T_C$-128,130,132, 134,125<br>$T_2$-20,40,33 | |
| $Tb_2NiMnO_6$ 122,146,155,167,188 | 1.04 | $P2_1/n$ | Ni-O = 1.99-2.04<br>Mn-O = 1.907-1.99 | 140 - 149.5 | 89.79 - 90.14 | $T_C$-111,110,113<br>$T_2$-12K,15,20,15 | |
| $Dy_2NiMnO_6$[140,150,151,153,177,188] | 1.027 | $P2_1/n$ | Ni-O =1.95-2.034<br>Mn-O = 21.91-1.924 | 146.92-148.5 | 90.23 - 90.25 | $T_C$ -93,105,95,101, 97,95,100<br>T' -21,6,20,50 | |
| $Ho_2NiMnO_6$ [121,122,153,186,194] | 1.015 | $P2_1/n$ | Ni-O= 1.99-2.04<br>Mn-O=1.91-1.99 | 145.6<br>146.3 | 89.72 - 90.21 | $T_C$-82,79,86,93,85 | |
| $Er_2NiMnO_6$ [153,155] | 1.004 | $P2_1/n$ | Ni-O = 2.04<br>Mn-O = 1.91 | 145.33 | 90.22 | $T_C$ - 84,74 | |
| $Tm_2NiMnO_6$ [155] | 0.994 | $P2_1/n$ | Ni-O = 2.044<br>Mn-O = 1.915 | 144.16 | 90.29 | $T_C$ - 62 | |
| $Yb_2NiMnO_6$[108] | 0.985 | *Pnma* | Ni/Mn -o = 1.96 | - | - | $T_C$ > 350 | |
| $Lu_2NiMnO_6$ 156,157,159 | 0.977 | $P2_1/n$ | Ni-O = 1.95<br>Mn-O = 1.99 | 142.5-145.5 | 90.32 - 90.69 | $T_C$ - 45,40<br>$T_C'$ -20 | |
| $Y_2NiMnO_6$ 122,145,162,177,195 | 1.02 | $P2_1/n$ | Ni-O = 2.018<br>Mn-O = 1.938 | 145.95<br>147.18 | 89.74, 90.141 | $T_C$ -79,81<br>T ' - 10 | |

Valence state of cations are $R^{3+}$, $Ni^{2+}$ and $Mn^{4+}$(small amounts of $Ni^{3+}$ and $Mn^{3+}$ is also seen in XPS analysis of some reports).
$T_C$ - the ferromagnetic curie temperature,
$T_C'$-The secondary transition due to antisite disorders
T'- the magnetic anomaly at low temperature

## Structural Properties of PNMO

PNMO double perovskite, which is a less explored member of the RNMO series, in which Mn and Ni are arranged in alternatively to produce rock salt type arrangement. $Pr^{3+}$ (ionic radii ~0.99 Å) occupies the A site and $Mn^{4+}$ (~0.53 Å), $Ni^{2+}$ (~0.69 Å) in alternate octahedral B' and B'' lattice sites in $A_2B'B''O_6$. It exhibits two types of crystal structure based on the Ni and Mn crystallographic site occupancies. First one is the orthorhombic crystal structure with the random arrangement of Ni and Mn. Second one is the monoclinic structure with Ni and Mn cations in alternating octahedral arrangement (Fig:7) [122,155,182] with lattice parameters $a \approx b \approx \sqrt{2a_p}$ [66,196]. Mn and Ni ions occupy the same Wycoff position 2*b* in the orthorhombic *Pnma* structure. In PNMO according to the following reaction, $Ni^{3+}+Mn^{3+} \leftrightharpoons Ni^{2+} + Mn^{4+}$ charge disproportionation leads to the formation of $Mn^{4+}$ and $Ni^{2+}$ [97,197].

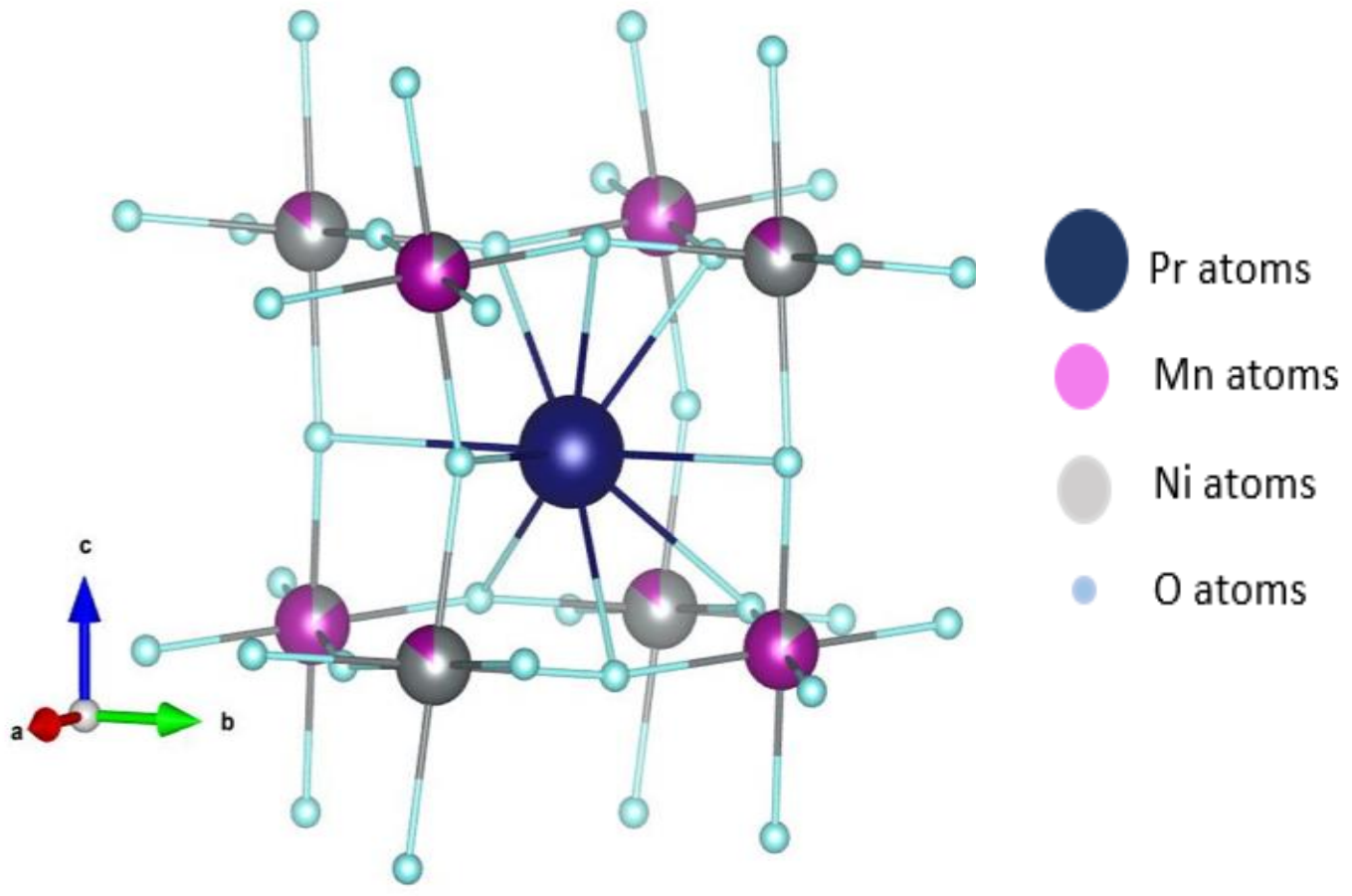

*Figure 3: Perovskite Unit of PNMO*

Various synthesize routes have been reported for PNMO material [4,26,31,87,88,120–122,152,166,168,176,181,198–200] such as modified nitrate decomposition route [120], glycine nitrate method [198], sol gel process [4,121], Hydrothermal route[200], Pechini method [167], solid state reaction route [26,88,122,152,166,168,199]. The ordering of Ni, Mn also depends on synthesis technique [197]. The cation composition and deviation of oxidation states of $Ni^{2+}$ and $Mn^{4+}$ are affected by the oxygen partial pressure of annealing atmosphere and they in turn govern their magnetic properties. The influences of these parameters on the dielectric properties have also been reported in the literature [201,202]

Determination of the pattern of B cations ordering is difficult with XRD analysis due to the resemblance in scattering factors. Neutron diffraction is an important tool to ensure the ordering of B'/B" sites. Rietveld refinement of XRD data well fitted with the single phase of PNMO with monoclinic *$P2_1/n$* space group [26,88,120–122,152,167,168,198–200]. Rietveld refinement of the XRD patterns of PNMO at room temperature is shown in figure :1 by Lekshmi *et al.*[167]

Bull *et al.* [120] indexed the peaks corresponding to an orthorhombic unit cell *Pnma.* While due to B cation ordering less symmetric monoclinic *$P2_1/n$* is observed through the neutron diffraction analysis. In the monoclinic symmetry $Ni^{2+}$ and $Mn^{4+}$ ions occupy 2*c* (0,1/2,0 ) and 2*d* (1/2,0,0) sites respectively and $Pr^{3+}$ and $O^{2-}$ ions are occupy at general lattice sites 4*e* (x, y, z) [203,204]. The structural parameters obtained from XRD are tabulated in Table 2.

| *Table 2: Lattice parameters, bond length, bond angle obtained from XRD* | | | |
|---|---|---|---|
| Reference | Lattice parameters | Bond length | Bond angle Ni-O-Mn |
| Zhang *et al.* [21] | a = 5.4339 Å, b = 5.4451Å, c =7.6910Å β = 89.497$^0$, V = 227.55 | Mn-O=1.892Å<br>Ni-O= 2.166 Å | 158.014$^0$ |
| Truong *et al.* [22] | a = 5.54 Å, b = 5.50 Å c = 7.70 Å | Mn-O=1.928Å<br>Ni-O = 2.029Å | 158.6$^0$ |
| Booth *et al.* [16] | a = 5.4453 Å, b = 5.4701Å, c = 7.696Å, β = 90.030$^0$, V = 229.257 | Mn-O=1.928Å<br>Ni-O= 2.029 Å | 158.6$^0$ |
| Balasubramanyam *et al.* [20] | a =5.4672Å b = 5.5362Å c= 7.7336 Å β = 89.88$^0$, V = 234.07 | Mn-O=1.93Å<br>Ni-O= 2.04 Å | - |
| Nazir *et al.*[15] | a =5.44 Å b = 5.47 Å c= 7.69 Å β = 90.01$^0$, V= 228.5 | Ni/Mn -O =1.97 Å | 157$^0$ |
| NeenuLakshmi *et al.*[30] | a = 5.4455, b= 5.4692, c = 7.6964, β =90.01$^0$, V = 229.22 | - | 155.6$^0$ |

From the table 2, it is obvious that there is a structural change in the PNMO monoclinic angle $\beta$ from 90$^0$ which is due to the lower ionic radius of $Pr^{3+}$ which in turn cause octahedral tilting for attaining stability. With maintaining the corner sharing connectivity, the $MnO_6$ and $NiO_6$ octahedra undergo tilting in order to overcome the instability arising due to the ionic radii mismatch of the A and B'/B" site cations. This type of distortion is common in compounds that include very small A cations to occupy the 12-coordination site. This results in the 8 coordination A site in the monoclinic RNMO samples [69]. The Ni/Mn-O bond lengths are limited to a narrow range. The Ni-O-Mn bond angles must distort with the octahedral tilts. In these tilted compounds the monoclinic $\beta$ angle decreases as the $r_R{}^{3+}$value decreases.

In parallel the decrease of the ideal 180° Ni-O-Mn bond angle also takes place.[72] This distortion leads to decrement of Ni-O-Mn bond angle from $180^0$.

In the notation developed by Glazer [205] for representing tilt system $a^0a^0a^0$ is the aristotype cubic structure with $Fm\underline{3}m$ space group symmetry [206]. There are two types of tilt system possible in this crystal structure depending on the axes through which the tilting occurs. By looking the crystal structure through monoclinic $P2_1/n$ [110, $\underline{1}$10] axes, successive octahedra undergo antiphase tilting down these axes. However, along the monoclinic $P2_1/n$ [002] axis, successive octahedra undergo an in-phase tilting [204]. In this tilt system equal magnitude of subsequent octahedral tilts with opposite direction about the pseudo cubic *a* and *b* axes can be viewed. Hence the representation for this tilt system is ($a^-a^-c^+$). While another tilt system in which in-phase tilting along [01$\underline{1}$], and an antiphase tilting along [011, 21$\underline{1}$] axes lead to ($a^+ b^- b^-$). Hence Glazer representation of the two octahedral tilting systems for the $P2_1/n$ monoclinic phase of PNMO is written as ($a^-a^-c^+$) and ($a^+ b^- b^-$) [9,207].

In PNMO films the ordered Mn/Ni site arrangement can be stabilized through moderately high temperatures and high oxygen pressure growth of the films. These films remain in short-range order regardless of growth conditions [199]. There is a presence of small amounts of randomness in configuration of Mn and Ni even in completely ordered monoclinic structure. These disorders are of different kinds based on the way it is distributed within the lattice. First one is the antisite (AS) defect, where Ni and Mn cations just interchange their sites. Slight volume of such AS defects is predicted to occur due to configurational entropy. antiphase boundary (APB), is another type of disorder which parts two ordered domains with inverted Ni and Mn site occupancies. This can be explained in other ways as building up of AS disorders in one place, forming the exact pattern of the ordered phase, but in the opposite direction [208]. This results in $Mn^{4+}$-$Mn^{4+}$ and $Ni^{2+}$-$Ni^{2+}$ anti-ferromagnetic interactions. When the antisite defects and random occupancies are high, then $Mn^{3+}$ and $Ni^{3+}$ regime forms, in which superexchange mechanism is through antiferromagnetic interactions. Consequently, secondary magnetic transition or occasionally a glassy state at low temperatures can happen. [97,121,122,174,197,199]. Furthermore, compared with the well-studied LNMO and NNMO compounds, the values of saturation moment in magnetization studies reveal that PNMO can be obtained with negligible anti-site defects [26].

**XPS Analysis of PNMO**

Pr 3*d* XPS spectrum and Mn and Ni 2*p* XPS spectrum in fig:2 Li *et al.*[168] reveals the presence of $Pr^{3+}$ valence states and the mixed valency of $Mn^{4+}/Mn^{3+}$ and $Ni^{2+}/Ni^{3+}$ in PNMO [168,198]. In the Pr 3*d* XPS spectrum, two broad peaks corresponding to Pr $3d_{3/2}$ at 954 eV and Pr $3d_{5/2}$ at 933 eV confirms the existence of $Pr^{3+}$ in PNMO material. Shake off Satellites is observed in the lower binding energy regime for this doublet[209]. Peaks corresponding to $Pr^{4+}$ are at 925.5 eV and 943.3 eV [209]. Since these peaks are absent in the XPS, spectra reveal that in PNMO Pr is at 3+ oxidation state. For Mn 2*p* spectra, the Mn $2p_{1/2}$ is assigned to peak at 652 eV and Mn $2p_{3/2}$ at 641 eV. For Ni 2*p* spectra the peak at 872 eV is of Ni $2p_{1/2}$ and at 854 eV is of Ni $2p_{3/2}$. The presence of binding energy peaks of $Mn^{3+}$, $Mn^{4+}$ in Mn 2*p* spectra ($Mn^{3+}$ $2p_{1/2}$ - 651.9 eV , $Mn^{3+}$ $2p_{3/2}$ -640.8 eV, $Mn^{4+}$ $2p_{1/2}$ - 653.8 eV , $Mn^{4+}$ $2p_{3/2}$ - 642.3 eV) and presence of $Ni^{2+}$ , $Ni^{3+}$ in Ni 2*p* spectra ($Ni^{2+}$ $2p_{1/2}$ - 871.64 $Ni^{2+}$ $2p_{3/2}$ - 853.96 eV, $Ni^{3+}$ $2p_{1/2}$ - 873.62 eV, $Ni^{3+}$ $2p_{3/2}$ - 855.55 eV ) confirms that Ni and Mn in PNMO are in mixed valence state [210]. Because of the presence of mixed states of $Mn^{4+}/Mn^{3+}$ and $Ni^{2+}/Ni^{3+}$, Bull *et al.* [120] reported small polar hopping conductivity of the sample at room temperature. Moreover, the percentage contribution of each valence state is calculated as 58.38%, 41.61%, 70.70%, 29.30% for $Mn^{3+}$, $Mn^{4+}$, $Ni^{2+}$, $Ni^{3+}$ respectively [168,198]. The B site cation disorder has an effect on the oxidation state of Mn and Ni. In ordered samples oxidation state combinations of $Ni^{2+}/Mn^{4+}$can occur, but more disordered samples exhibit $Ni^{3+}/Mn^{3+}$ states [115,117,119]. For example, Asai *et al.* [181,211] have done NMR spectroscopic analysis of LNMO in which the ordered compound had $Mn^{4+}$ state, while $Mn^{3+}$ state is shown by Mn ions in the Antisite defect.

Nazir *et al.* [121] have done EXAFS analysis of RNMO (R=La to Ho) to estimate the local environmental changes in the (Mn/Ni) $O_6$ octahedra and Mn/Ni-O bond lengths. A main peak at ~ 1.35 Å appears in the EXAFS spectra. This matches the first coordination shell of Mn to the six oxygen atoms. A projecting peak around 1.44 Å indicates the Ni-O coordination cell.

**Raman Analysis with PNMO**

Raman spectroscopy is useful in examining the crystalline structure, spin−phonon coupling, disorder of cations, presence of impurity phases and local/dynamical lattice distortion in the double perovskite structure [34,212]. No Raman active vibrational modes are observed for the ideal cubic perovskite structure [120]. Rotation/tilting of octahedra, displacements of the A or B cations, or B site ordering leads to alterations from cubic symmetry. This results in the symmetry reduction and Raman active modes will originate at the Brillouin zone centre. In Raman spectra the number of peaks and the relative intensity of these peaks is associated with extent of alteration from the aristotype cubic structure [120]. Raman spectroscopy is a non-destructive tool to differentiate between the monoclinic $P2_1/n$ and the orthorhombic *Pbnm* perovskite structure; this is probable due to selection rules and the different symmetry of the corresponding Raman modes [213,214].

Singh [199] have analysed the micro-Raman spectra of PNMO films in XX, XY, X’X', X'Y' scattering configurations. In this X and Y represent the horizontal and vertical polarization directions, while X', Y' represent diagonal directions to X and Y. The significant intensity reduction in the XY mode and absence of these modes in the X'Y' configuration proposes the monoclinic and/or orthorhombic symmetry of PNMO material [215,216]. In some reports these Raman spectroscopic arrangements are represented as HH (XX) and HV(XY). corresponding to parallel and perpendicular directions of configurations of polarized Raman spectroscopy arrangement. Experimentally, for PNMO monoclinic ($P2_1/n$) phase two sharp peaks are detected in two polarization configurations (HH, HV) is shown in Truong *et al*. [87] in figure 3 and *Mayer et al*.[217] in figure 2.

Depending on the polarization arrangement there is variation in the relative intensity of Raman peaks. [199] In the Raman spectrum of PNMO two strong modes around 650 $cm^{-1}$ ($A_g$) and 500 $cm^{-1}$($B_g$) were observed [87,120,121,199]. These vibrational modes are due to the symmetric stretching of $(Ni/Mn)O_6$ octahedra (breathing mode) [213,217,218] and mixed kind of antisymmetric stretching and bending motions (mixed mode) [121,217].

The mode at 650 $cm^{-1}$ is having $A_g$ symmetry, which is acceptable in HH configuration. Relatively less intensity of this mode is found for crossed HV scattering arrangements. The phonon mode at 500 $cm^{-1}$ exhibits $B_g$ symmetry with high intensity at HV configuration and comparatively less intensity at HH configuration [213,218–220]. The crystallinity of the grown PNMO films can be indicated by the modes around ~1200–1350$cm^{-1}$. These additional modes are the second order overtones of the $A_g$ and $A_g$/ $B_g$ mode combination [220]. Around the low frequency region (~300 $cm^{-1}$) feeble stretching modes indicate $P2_1/n$ monoclinic symmetry. For the confirmation of the pure phase of the samples there is absence of Raman peak around 377 $cm^{-1}$ corresponding to most of the rare earth oxides [221].

The broadness in the Raman modes is due to the antisite disorders. The disordered Ni and Mn sites contribute different Ni-O and Mn-O stretching vibrations. These unresolved vibrations having similar frequency in the B'/B" sites also added up in the band of Raman modes that increase the bandwidth [151]. As sample cools, softening of $A_g$ mode is observed, which is initiated at the transition temperature and continues as progressively cools the sample [217]. Truong *et al.* [87] in figure:4 express the shift in the $A_g$ mode and $B_g$ mode in Raman spectrum with temperature .This softening may be due to two reasons, first one is that the interconnection between Ni/Mn–O exchange interaction and force constant may be negligible. Second reason is that the variation of the exchange interaction is feeble by itself due to temperature change in PNMO [87]. As progressively cool the samples an increase in intensity of the phonon excitations and a decrease in their linewidth is observed. This indicates the low temperature reduction of phonon scattering and improved Raman tensors due to reinforced orbital polarization.

The dependence of temperature on the frequency of the phonons have two main origins; (i) anharmonicity [222], and (ii) spin–phonon coupling [223]. For anharmonic term, the position of the phonon obeys Balkanski's model as the temperature changes [222]. That considers the anharmonicity contributions with the relation,

$$w(T) = w_0 - C\left[1 + \frac{2}{\left(e^{\frac{\hbar w}{KT}} - 1\right)}\right]$$

with C and $\omega_o$ being fitting parameters.

Spin phonon coupling is related to phonon renormalization as a consequence of magnetic ordering. This is because of the coupling between magnetic ordering and lattice [110,215,224]. The contribution of spin–phonon coupling $\left(\Delta w_{ph}\right)_K$ to position change of the kth phonon is expressed as [223]

$$\left(\Delta w_{ph}\right)_K = \sum_{i,j>i} J_{ij} < s_i . s_i >$$

$J_{ij}$ represent the super exchange integral and $< s_i . s_j >$ refers to the spin correlation function. Here to arrive at this relation it is considered that the magnetostriction effects and electronic states renormalization are absent and also considered only the first-neighbour interactions.

Also, a molecular field approximation, can be considered in the case of the stretching phonon, as

$$\left(\Delta w_{ph}\right)_K \propto \left(\frac{M(T)}{M_0}\right)^2$$

M (T) is the magnetization at a particular temperature T, and $M_0$ is the maximum magnetization obtained.[110,223]

When the rare earth ion of RNMO has changed with another rare earth ion, the shifting of Raman peaks is observed. The decrement of $r_R^{3+}$ shows a shifting of Ag mode towards lower wavenumbers is observed. This is due to the decrease of force constant of the perovskite lattice with increase in the Ni/Mn -O bond length. This results in symmetric stretching mode to shift to higher wavenumber. Hence decrease in the A cation radii leads to the reduction in the superexchange interaction and also spin phonon coupling [217]. The spin phonon coupling inter relates the magnetic and phonon properties. This kind of behaviour is also shown by rare earth manganites $RMnO_3$ [225,226].

Truong *et al.* [87] reported the beginning of peak softening is around 100 K for polycrystalline PNMO samples, which is far below the $T_C$ of PNMO at 228 K (Fig:10). Compared with the values for the softening of $A_g$ mode in LNMO, which is around 7 $cm^{-1}$ from $T_C$ to low temperatures (∼10 K), the softening of PNMO is small ∼1.5 $cm^{-1}$[213,216,218,227]. These results predict that the influence of force constant on the variation of Ni/Mn and O ions exchange interaction is small in PNMO [226]. Another study of PNMO thin films, around the ferromagnetic transition temperature 214 K the softening of peak ($A_g$) is observed [199]. Temperature dependence of the $A_g$ mode in epitaxial film and in polycrystalline bulk sample are pictured by Singh *et al.* in figure:4[199]. This softening is only about 2 $cm^{-1}$, which indicates a weak spin phonon coupling.

Compared with the Ni, Mn system (PNMO) for the Co, Mn system (PCMO) The B–O stretching modes observed about 20-30 $cm^{-1}$ higher wavenumber [120,217]. Clarifications on this regard is that in the Ni-Mn system, the Ni/Mn–O and Pr–O bonds have a better extent of overlapping compared with the Co-Mn scheme. The large extent of orbital overlapping observed in the Ni-Mn system reduces the bond length. Also, the bond strength can be changed by varying transition metal charge density [120,217]. Meyer *et al.* [217] reported $A_g$ mode softening in PNMO is initiated at $T_C$ around 220K and further increased by successive cooling [217]. The weak impact of $Co^{2+}$ and $Ni^{2+}$ cations on the spin-phonon coupling is indicated by slight change in the spin-phonon coupling strength, for the same A-site cation. In general, the $R_2CoMnO_6$ (RCMO) film series shows slightly larger values of spin phonon coupling strength (only exception R = La) than the RNMO film series [217]. As the cation size of $Co^{2+}$ and $Ni^{2+}$ is similar [180] and the lattice parameters are also comparable, apart from the extent of overlapping, other factors like the different spin contributions of HS–$Co^{2+}$/$Ni^{2+}$, and their impact on the bonds, force constants, and phonon properties may involve in this regard. The impact of B site ordering and the spin phonon coupling in double perovskite is investigated in the study of Gd (Co, Mn) $O_3$ and its spin phonon coupling observations [224].

**Magnetic properties**

Magnetic properties of PNMO are related to the Ni-O-Mn bond angle and the covalency/ionicity of Ni/Mn-O bond. Thermo magnetization curves FC and ZFC are used to study the magnetic properties of these systems. In Field cooled cooling (FCC) approach the sample is cooled from a higher temperature along with an application of magnetic field and throughout the procedure recording the magnetization. The temperature was then increased with applied field and measured the magnetisation (FCW). However, in the ZFC protocol the sample cooled in the absence of a magnetic field. Then the magnetization is measured by increasing the temperature along with the magnetic field. Magnetization plots in zero field cooled (ZFC) and field-cooled (FC) protocols show a divergence between FC and ZFC at low temperature. Lekshmi *et al.*[167] in figure :2 show the thermo magnetization field-cooled (FC) and zero-field-cooled (ZFC) curve of PNMO and dM/dT versus T. The applied magnetic field is 500 Oe. The divergence initiated near the transition temperature. This indicates competing magnetic interactions like spin glass or cluster glass [31,121,167]. Due to magneto crystalline anisotropy resulting from the spin-orbit coupling between the Ni-Mn network and Pr ions [88,122,166]. Variation of dm/dT with T and it shows only one magnetic transition for PNMO. The lower magnetic transitions occurred due to the presence of antiphase boundaries that are absent in the material. This confirms the ordered arrangement of PNMO compared to many other RNMO samples.

Near room temperature, the PNMO sample is paramagnetic. While, with decreasing temperature a magnetic transition is observed around 200 K. The critical temperature varies from 208K to 255K in different reports [26,88,121,122,166,198–200]. Paramagnetic curie temperature obtained from Curie Weiss plot has values 218 to 226 K [88,121,122]. Observed effective paramagnetic moment (Results from fitting to the Curie–Weiss plot by the equation $\mu_{eff} = 2.828 \sqrt{c}$ )[195] ranges from 6.124 to 8.41 $\mu_B$/f. u. [121,122,200]. Effective magnetic moments can obtain using the equation [121,155] :

$$\mu_{eff}(calc) = [2\mu_B(Pr^{3+})^2 + \mu_B(Ni^{2+})^2 + \mu_B(Mn^{4+})^2]^{1/2}$$

Reported $\mu_{eff}$ (Calc) values are 6.32 to 7.14 $\mu_B$/f. u [121,122,200]. Measured saturation moment (M) has values from 4.92 to 5 $\mu_B$/f. u [26,121,122,199,200]. From the M vs 1/H plot The extrapolation to 1/H =0 will give the value of saturation moment ($M_{extrapolated}$) [121,122] reported as 5.48 to 6.69 $\mu_B$/f. u [121,122,200]. Theoretical saturation moment Ms = [$2g_J$ + 5.0] μB/f. u [121], where $g_J$ is the rare-earth saturated moment, and 5 μB/f. u is the effective magnetic moment of the Ni-Mn system. $M_s$ is in the range 11.4 to 13.7 $\mu_B$/f. u [121,122,200]

| **Effective magnetic moment $\mu_{eff}$ calculated for possible spin state configurations of Ni and Mn in PNMO** | |
|---|---|
| Spin systems | $\mu_{eff}$ (calc) = $[\,2(\mu_{Pr})^2 + (\mu_{Ni})^2 + (\mu_{Mn})^2]^{1/2}$ |
| $Mn^{4+}$ (S = 3/2), $Ni^{2+}$ (S = 1) | 6.971 |
| $Mn^{3+}$ (S = 2), $Ni^{3+}$ (S = 3/2) | 8.035 |
| $Mn^{3+}$ (S = 2), $Ni^{3+}$ (S = 1/2) | 7.252 |
| $Mn^{2+}$ (S = 2), $Ni^{4+}$ (S = 2) | 8.578 |
| $Mn^{2+}$ (S = 2), $Ni^{4+}$ (S = 1) | 7.589 |
| $Mn^{2+}$ (S = 2), $Ni^{4+}$ (S = 0) | 7.042 |
| $\mu_{Ni}$ / $\mu_{Mn}$ (spin only systems) = $[g\,S(S+1)]^{1/2}$; g = 2 for Mn and Ni<br>$\mu_{pr} = [\,g_{Pr}^2\,J(J+1)]^{1/2} = 3.578$<br>$g_{Pr} = \frac{3}{2} + \frac{S(S+1)-L(L+1)}{2J(J+1)} = 0.8$ (S = 1, L = 5, J = 4) | |

In the table 4 the effective magnetic moment of all the possible oxidation states of Ni and Mn are calculated for comparison.[228,229] From the values the experimentally determined values are in correlation with the oxidation state of ordered $Mn^{4+}$ and $Ni^{2+}$ and $Mn^{3+}$ and low spin $Ni^{3+}$. The $Ni^{4+}$ (s=0) and $Mn^{2+}$ (s=2) also have values of magnetic moment in the range of experimentally obtained results. But the zero spin of $Ni^{4+}$ can enable the choice of this pair of ions in the PNMO material. The studies with NMR and XAS [115–117] show the actual valence states are $Ni^{2+}$ and $Mn^{4+}$ . The theoretical and experimental correlation of these values is due to the consideration of spin only systems $Mn^{4+}$ and $Ni^{2+}$ with $\mu_{eff}$ ($Mn^{4+}/Ni^{2+} = g\sqrt{s(s+1)}\mu_B$) and $Pr^{3+}$ with $\mu_{eff}$ ($Pr^{3+}$) = $g\sqrt{J(J+1)}\mu_B$) since J is considered a good quantum number for lanthanides. The difference in the effective paramagnetic moment value for experiment and theoretical is associated with the crystal field splitting. In the presence of oxygen ligands on the $MnO_6$ and $NiO_6$ octahedra the degenerated d orbitals will split into $t_{2g}$ and $e_g$ states. The $e_g$ state is having the higher energy in the spherically symmetric environment.[186] The filling of electrons in these orbitals is based on Hund's rule as a consequence of the competition between the coulomb energy and crystal field energy. In certain systems the crystal field splitting is stronger than the spin orbit interaction. This is called orbital quenching.[230] For 4f electrons, the orbitals are near to the nucleus and inside 6s and 5p. Hence the crystal field is shielded by the 6s and 5p electrons. For 3d electrons the shielding of 4s electrons is not as strong as that for 4f electrons. This confirms the prominent crystal field splitting of these compounds which leads to the orbital quenching and becomes spin only systems. The critical temperature of PNMO is less than LNMO and greater than other members of RNMO series which have ionic radius less than Pr. Which can be explained on the basis of the change in superexchange angle discussed before.

Antiphase boundaries formed due to the antisite defects at Ni/Mn site leads to a negative ZFCM(T) in PNMO in the study done by Nazir *et al.*[121] ferromagnetic coupling of short-ranged order forms among $Ni^{2+}$ and $Mn^{4+}$ cations. While in the antiphase boundaries coupling is antiferromagnetic in nature. It is impossible to align these opposite spins in line with the direction of magnetic field at relatively lower magnetic field strength in the ZFC mode in anti-phase boundary. As temperature is increased the antiparallel or tilted spins get weakened [121]. The thermo magnetization curve obtained from the field cooled protocol shows a rapid increase in magnetization $T_c$ which is due to the ferromagnetic ordering of $Ni^{2+}$ and $Mn^{4+}$ sub lattice. For higher temperature this upturn moves to higher temperature, which is the common feature of ferromagnetic interaction [188].

Troung *et al.* [87] and Harisankar *et al.* [31] reported a secondary magnetic transition around 126K for PNMO thin films, which arises due to the presence of anti-phase boundaries. Ali *et al.* [188] theoretically reported for PNMO samples around 27 K there is a downturn in magnetisation at 27K in low magnetic fields. This downturn in the magnetic field can suppress and change to an upturn at larger magnetic fields [187,188]. There is a slight disagreement with the experimentally obtained results for PNMO. There is no downturn in $M_Z$ at low temperature in the low field data of PNMO. Magnetization ($M_Z$) as a function of temperature (T) at various strength of external fields the for the experimental and theoretical (Monte carlo results) are explained with figure:1 and figure:5 by Ali *et al.*[188] This may be due to smaller spin orbit coupling in PNMO that can be easily overcome even by a small magnetic field. The experimental findings are supported via Monte Carlo simulations and mean field analysis on the two-sublattice Heisenberg model [188] for other rare earth ions R=Nd, Sm, Dy Gd, Tb. Rare earth ions magnetic moments appear only at low temperature characteristically below 30 K [179]. In some reports this secondary magnetic transition below 30 K is seem to be the result of antiferromagnetic coupling between R ion (Pr, Nd, Sm) and Ni-Mn sublattice and ferromagnetic for R = Gd, Tb,

Dy [182,191,231]. However recent DFT analysis on $Nd_2NiMnO_6$ reveals that magnetic spin between Nd spin and Ni-Mn sublattice is ferromagnetic in nature. Hence Ali *et al.* [188] Explain this secondary transition as, for less than half filled 4f shells the orbital moment of the R ion is oppositely aligned to spin moment and is larger in magnitude than spin moment. Even though the spin is coupled ferromagnetically to the Ni-Mn sublattice the orbital moment is opposite to the spin moment, resulting in a downturn in magnetization. At large enough magnetic fields, it becomes energetically favourable for the net effective magnetic moment to align with the field, which leads to a switching of the magnetization anomaly at 27K from a downturn to an upturn as the field is increased.

Nevertheless, many of the reports [26,121,122,167,199,200] establish single magnetic transition in PNMO which arises due to the $Ni^{2+}$-O-$Mn^{4+}$ superexchange interaction and also absence of the secondary transitions is an implication of less disordered samples.

In the $1/\chi$ vs. T the high temperature region data are fitted perfectly with a Curie – Weiss (CW) law with positive $\theta$ values. This is an indication of the supremacy of the ferromagnetic relationship of spins in this system [121,122,166,167,200]. Reports on saturation magnetization reveal the ideal value $5\mu_B$ for complete ordered samples [122,200]. The slightly lower values than the $5\mu_B$ indicate the presence of a small amount of antisite disorders [26,166,199]. All paramagnetic temperatures obtained from the Curie- Weiss plot is positive. That proposes the predominant exchange interactions as ferromagnetic.

Observation of a significant Magnetic caloric effect (MCE) in double perovskite RNMO opens up possibilities for using the high degree of flexibility available in these oxides to enhance the effect further. Properties such as Superexchange interaction, low value of coercive field and high saturation magnetic moment for rare earths with high magnetic moment are vital for magnetocaloric refrigerant.

Based on thermodynamic theory, for a range of magnetic field $\Delta H$ the change in temperature and the corresponding magnetic entropy change ($-\Delta S_M$) is expressed as,

$$\Delta s_M(H,T) = \int_0^H \left(\frac{dM}{dT}\right) \ \left(\frac{dM}{dT}\right)_{H'} \ dH'$$

Relative cooling power (RCP) is used to quantify the usefulness of MCE materials which is defined as [88,188]

$$\text{RCP} = -\Delta S_M(T,H) \times \delta T_{FWHM}$$

Ali *et al.* [188] recently published a paper in which comprehensive experimental and theoretical study of magnetization and magnetic caloric effect of RNMO materials was reported. Distinct behaviour depending on the choice of R is observed at low temperature. This low temperature anomalies in magnetization manifest in the change in the entropy -ΔS whose sign depends on choice of R. Magnetic entropy changes versus temperature at different applied magnetic fields up to 8T for PNMO shows two prominent features (Fig:13). First one is at the ferromagnetic transition temperature $T_c$ and second one at lower temperature ($T_2$) [188]. The value of $-\Delta S_M$ is positive across $T_C$ and its magnitude increases with field, which is consistent with the ferromagnetic ordering of the Ni-Mn sublattice at $T_C$. When the temperature is lowered $-\Delta S_M$ monotonically decreases below $T_C$. On increasing the magnetic field this decrease can be suppressed and a positive peak or anomaly appears at low temperature and high fields. This can be realised as the opposite alignment of the total effective moment of the R ions with the Ni-Mn sublattice below $T_C$ at low fields. So, the magnetic entropy is high. At higher fields, the R moment switches and aligns with the Ni-Mn sublattice and also the applied magnetic field, resulting in a low magnetic entropy like in the case of the ferromagnetic transition at $T_C$. Using mean field approximation and Monte Carlo simulations on the minimal spin model the key features of the experimental results are verified. The anti-alignment of Pr with Ni-Mn sub lattice at low temperature leads to $-\Delta S_M < 0$ (not observed experimentally). This comparison of experimental and theoretical results is shown in figure:3 and figure:7 by Ali *et al.*[188] At sufficiently high magnetic fields this external field can enforce the alignment and hence the inverse magnetocaloric effect can change to magneto caloric effect. The change in entropy with the Temperature graph also does not match with the experimental results.

**Electrical properties**

The electronic band gap of RNMO samples depends on the orbital overlap of the different elements involved in the crystal structure. The bond angle and bond length can be used to estimate the bandwidth [232,233]. The electrical resistivity and the band gap in these perovskites [234] mainly administered by the B–O sublattice. orbital overlaps along the distorted cube edges lead to the formation of bands. the geometrical factors; super exchange angle, and the characteristics of TM ions at B'/B" sites determine the magnitude of the orbital overlap. As superexchange angle decreases orbital overlapping reduces, consequently the hopping of electrons between TM ions via superexchange interaction decreases [95]. The decrease in the orbital overlap makes the conduction band and valence band more separated in energy. Hence there is a possibility of band gap increase with $r_R^{3+}$ decrement, and an increased tilting angle of octahedral [120,121,176]. However, there are contradictory reports which claim the RNMO based double perovskites band gap remains largely unaffected due the charge transference from oxygen to $Mn^{4+}$ and $Ni^{2+}$ [26]. According to this concept, electron hopping integrals and orbital band width is not influenced by the Mn-O-Ni bond angles and the Mn/Ni-O bond lengths in the material. However, variation in super-exchange strength changes $T_C$ [26]. This is manifested from high pressure analysis on LNMO that holds the ferromagnetic character, with a minor disparity in $T_C$ and magnetic moment decrement even under 30 GPa pressure [235]. Polycrystalline materials show orders of magnitude difference in resistivity due to the grain boundaries. That is some of the insulating compounds are in fact poor metals and which show poor resistivity due to grain boundaries [236–238]. An example for that is $La_2NiCoO_6$ [239] .

Balasubramanyam *et.al.* [26] from the O1*s* XAS edge spectrum of PNMO obtained in the total electron yield mode, determines the band gap energy. For determination of $E_F$, spectrum of $LaNiO_3$, which is a metallic compound, and $LaMnO_3$, with a known band gap is used. At the rising edge of $LaNiO_3$ the position of $E_F$ was fixed. Around 1.1 eV above $E_F$ there is a rise in spectra for $LaMnO_3$. This can be considered as the band gap of $LaMnO_3$. From optical conductivity measurements the bandgap of 1.2 eV has been obtained [240]. In the same method the bandgap of PNMO is calculated as 0.9 eV. This method of band gap determination from O1*s* spectrum is not reliable because of the effects of O1*s* core hole [26]. However, an understanding of the range of the band gap values can be obtained. The value obtained from resistivity measurements is around 0.65 eV and a greater bandgap value is obtained as 0.75 eV through charge transfer studies. The density of states analysis shows a band gap of 1.2 eV for U-J >2 eV then further constant for U-J = 8 eV [26]. Density of states analysis reveals the P-d type band gap of PNMO material. From four probe measurements the band gap is obtained as 1 eV [120]. From UV Visible absorption spectroscopy using tauc plot the band gap energy is reported as 1.42 -1.5 eV [121] . Antisite disorders that are insulating regions are having an impact on experimental band gaps. The voids created while pressing the powders and the intrinsic effects of grain boundary leads to the changes in the exact result of resistivity. Also due to the addition of carrier production effects which are temperature dependent results in the alteration of resistivity and band gap values [60].

**Dielectric properties**

The two parameters of dielectric material, the relative dielectric constant and tanδ of PNMO with various temperatures, are shown in figure :3 Lekshmi *et al.*[167] The complex dielectric constant of a material in the AC electric field can be characterized by: [16]

$$\varepsilon^* = \varepsilon' + i\varepsilon''$$

where ε′ is the real part representing the energy storage (dielectric constant) and ε″ the imaginary part of complex permittivity, the loss of energy (tanδ) during each cycle of electric field.

Study on dielectric properties on RNMO samples done by Booth *et al.* [122] reveals low dielectric constants (ε = 15–25) with no dispersion suggest that there is perhaps no microwave magnetocapacitance effect in these materials. Particularly in PNMO having ε is 23 and tanδ is 0.013 [122]. Small value of dielectric constants and loss tangent and high value of resistivity ($10^7$ V-cm at 300 K) of RNMO oxides are consistent with the ordered double perovskite structure [241]. Lekshmi *et al.* [167] conducted the magneto dielectric study on PNMO material in which the frequency dependent ε' (T) and tan δ (T) plot, a relaxor-like dielectric behavior is observed (Fig:14). $T_{max}$ is the temperature at which tan δ reaches its maximum. As the measuring frequency increases the value of $T_{max}$ shifts to higher temperature. Double relaxations are observed in ε' (T) plots [167]. By using Koops-like model the experimental results can be explained which assumes this material as a multilayer capacitor [242]. Hopping mechanism of electron among Ni /Mn ions is the reason behind the dielectric relaxation arising from the grains? From the results, the effect of a 'colossal dielectric constant' is obtained. This is due to the fact that in addition to the relaxation due to electron hopping, at higher temperature the grain boundaries reach 'conduction relaxation' state as well. At a higher temperature the increase of ε' will occur because of the fact that the necessary condition for a rapid relaxation time for higher frequencies. Increases in $T_{max}$ and $E_{a1}$ with decrement in the radius of rare-earth ion ($r_R^{3+}$) obviously point out the electron hopping mechanism that is further suppressed with decrement in $r_R^{3+}$ [122,182]. The regular decrease of $T_{max}$ and $E_{a1}$ with the $r_R^{3+}$ indicates that the magnetic and dielectric properties are related with the crystal structure of PNMO. The Ni-O-Mn bond angle decreases with decrease in $r_R^{3+}$. This in turn results in the decrease of overlapping of the orbital. That reduces the hopping of electrons between Ni and Mn. Hence results in the decrease in $T_c$ and increase in $T_{max}$ and $Ea_1$ for dielectric relaxation. It shows interconnection between the magnetic properties with dielectric properties of RNMO. This correlation is an indication of a spin phonon coupling which is really important for spintronics applications [167].

**A site modification in PNMO**

Rare earth ions have a great role in modifying the superexchange interaction and consequently the magnetic and electrical properties. Modification of the A site having $Pr^{3+}$ion with other elements will lead to the alteration of various properties of PNMO material. One of the studies for Ca doping in PNMO is used for the electrochemical applications. XPS analysis of the Ca doped PNMO which forms $Pr_{2-x}Ca_xNiMnO_{6-\delta}$ ($x$ = 0.0 ~ 0.3) ($PCNMO_x$) were studied[198]. The presence of $Ca^{2+}$ valence state was obtained. The occupancy of the $Ca^{2+}$ in $Pr^{3+}$ site will increase the defect states which lead to the increase in electrical conductivity of these samples. Glycine nitrate method is used to prepare $PCNMO_x$ double perovskite with monoclinic ($P2_1/n$) crystal structure. As the $Ca^{2+}$ doping increases the oxygen vacancies in the lattice also increases. This in turn results in the increment in the unit cell volume and thermal expansion coefficient. $\delta$ value of oxygen vacancies in $PNMO_x$ for x=0 is determined from the XPS spectra as 0.08. This gradually increases to 0.22 for x=0.3 [198]. Another modification is with $La^{3+}$ ions to form $LaPrNiMnO_6$. It has the same monoclinic structure. Since $La^{3+}$ is having larger ionic radii than $Pr^{3+}$ results in the increase in curie temperature compared to PNMO. It shows a magneto dielectric coupling in these samples.[166] Tang *et al.* [4] investigated the photocatalytic application of $LaPrNiMnO_6$. UV Visible analysis of Mesoporous $LaPrNiMnO_6$ shows the light absorption in the entire range of 200-700 nm indicates the effective absorption in the whole range and can be a good photocatalyst. This mesoporous structure has high oxygen content and better redox capacity. The activity of the catalytic oxidation of these samples also shows good results for the degradation of volatile organic pollutants.

**Theoretical studies on PNMO**

Zhao et al. [176] using first principal calculation studied the effects of chemical and hydrostatic pressures on structural, magnetic, and electronic properties of RNMO (R= Ce to Er) double perovskites. The structure correlation of many properties is explained on a theoretical basis. For both $Ni^{2+}$ and $Mn^{4+}$ the study has taken the Hubbard U value as 3 eV, which gives good correlation with the experimental results. For all the RNMO samples (R= Ce to Er) the value of mean Ni - O is larger than mean Mn-O bond length. The Ni-O bond lengths within these RNMO samples differ by less than 2.2% and the Mn-O bond lengths differ by less than 1.1% in comparison with experiments [122]. In the study they have tried to investigate whether applying hydrostatic pressure on SNMO has the same effect as the chemical pressure in RNMO samples on the structural and magnetic properties. The effect of chemical pressure (change in $r_R^{3+}$) we have discussed in the comparison section of RNMO samples in this paper. While considering the hydrostatic pressure the lattice parameters, mean Ni- O and Mn-O bond lengths obtained as decreasing with increasing hydrostatic pressure. The monoclinic angle, antipolar displacement and the octahedral tilting angle has little effect on the hydrostatic pressure. The electronic band gap and the magnetic curie temperature slightly decreases as the hydrostatic pressure increases. The study proposes new predictions in anticipation of experimental validation. In contrast to the general concept for rare-earth perovskites, response of many physical quantities is in dissimilar manner to chemical and hydrostatic pressures which needs to be verified experimentally. Also, the effect of these pressures on antipolar displacements need experimental confirmation. Which need additional explanation for the effects of rare earth ion in inducing and tuning of electrical polarization of LNMO/RNMO superlattices [133]. The effect of chemical and hydrostatic pressure on RNMO sample in Lattice parameters, Monoclinic angle, Bond length, tilting angle, Ni-O-Mn angle and Antipolar displacement are shown in figure:1 and figure:3 in the theoretical paper of Zhao *et al.*[176]

2*p* XAS spectra indicate the presence of $Mn^{4+}$ and $Ni^{2+}$ states for PNMO. Charge transfer energies for Ni and Mn ions are calculated from the XPS spectra of each ion. PNMO obtained a higher electron count in comparison with the ionic values for the ground states of $Ni^{2+}$ and $Mn^{4+}$ ions. According to the Zannen-Sawatzy-Allen phase diagram [243]. PNMO is a change transfer insulator with p-d type band gap and an intermediate covalent character[26] . Ali *et al.*[188] reported recently that the choice of R, in addition to affecting the value of $T_C$, also influences the magnetization and consequently the MCE behavior as inferred from the nature of low-temperature anomalies for different R. In the paper they propose a simple explanation for this behavior in terms of a Heisenberg model that takes into account the coupling of the spin of rare earth ion with the spin of Ni-Mn network, as well as the local spin-orbit coupling on R site. The magnetic behavior at low temperatures in the double perovskites containing f block elements is controlled by spin-orbit coupling. More than or less than half filled 4f orbitals decide the nature of spin-orbit coupling to be ferromagnetic or antiferro-magnetic. The weak ferromagnetic coupling J between the rare earth ion sublattice and Ni-Mn sublattice and antiferromagnetic spin-orbit coupling ($\lambda$) in PNMO is responsible for the downturn in magnetization at low temperature. These calculations are confirmed with the Monte Carlo simulations [188]. $Pr_2NiMnO_6/La_2NiMnO_6$ superlattice ground state structural electric and magnetic properties are investigated using first principal calculations [244]. The superlattice is having a ferromagnetic ordering. Consistent with the antipolar displacement of Pr and La ions a ferroelectric polarization is predicted.

In the field of magnetic materials, the study of phase transitions is important. The magnetic transitions can be classified into two as first order and the second order [245,246]. The detailed study of critical behaviour near the magnetic phase transition provides the knowledge about the microscopic interactions in the second order materials [247–254]. These critical properties can be analysed with four major theoretical models in second order transitions [255–257]. The magnetic transitions in most of the RNMO and RCMO compounds around paramagnetic-ferromagnetic (PM-FM) transitions are determined to be of second order transitions [152,258–262]. Su *et al.* have studied the magnetic critical properties of the second-order PM-FM transition of RNMO (RE = Eu and Dy). The study reveals the critical behaviour of $Eu_2NiMnO_6$ are in between the mean-field and (3D) Heisenberg interaction models. For $Dy_2NiMnO_6$ the critical behaviour cannot be explained by any known theoretical model. This may be due to the nonexistence of long-range order in DNMO [140]. The critical parameters for RNMO (R = Y, La) are very close to the 3D-Heisenberg theoretical model with a short-range FM order [263,264].

## Applications

### Electro chemical property and energy storage application

PNMO has been proved as a good candidate as cathode for Intermediate temperature Solid oxide Fuel cells (IT-SOFC). It exhibits high chemical stability, thermal expansion characteristics, with GDC (Gd doped $CeO_2$ – $Ce_{0.9}Gd_{0.1}O_{1.95}$) as electrolyte. Also shows promising electrochemical performance around $700^0$C in air. The electrical conductivity of PNMO is obtained as very low as approximately 3 S $cm^{-1}$ at $800^0$ C. By the optimization of the electrode microstructure and formation of composite electrodes the cathode efficiency can enhance.[168] Sun *et al.* [198] in figure :7 described the schematic of the ORR reaction mechanism and the impedance analysis plot for the PNMO cathode.

. Ca doping in PNMO and consequent electrochemical performance of $Pr_{2-x}Ca_xNiMnO_{6-\delta}$ were studied using AC impedance spectroscopic analysis at $700^0$C in air. The reaction mechanism of the cathode is not changed with calcium doping. The increment in the oxygen vacancies due to doping of Ca and corresponding enhancement in the conductivity results in the better electrochemical performance. The possible application of Ca doped PNMO for SOFCs has been verified with I-V and I-P curves of single cell. The long-term stability of this particular cathode material is obtained from the constant value of the voltage and power density throughout the measurement [198]. Recently supercapacitor application of GNMO has been reported [142]. This can be extended to PNMO material as well.

**Photocatalysis**

The higher oxygen content and redox capacity of the mesoporous double-perovskite catalysts $LaPrNiMnO_6$ can be used in the application of photothermal synergistic degradation of gaseous toluene. Schematic of the photocatalysis process involved in the material is shown in fig:18. Larger specific surface area and consequent increase in the reactivity of the catalyst are the important features of mesoporous structure. $LaPrNiMnO_6$ possessed high capacity for the adsorption of oxygen content and also absorption of light is obtained from UV vis and XPS spectra analysis. Which results in the occurrence of catalytic oxidation in Mars–van Krevelen redox cycle mechanism [265]. $LaPrNiMnO_6$ exhibited better catalytic activity due to the lower activation energy in the reaction kinetics study of this material. This property opens the door into the wide application of RNMO and modified RNMO samples in the field of volatile organic pollutant degradation [4]. Possible $C_7H_8$ degradation mechanism of $LaPrMnNiO_6$ underphotothermal conditions schematically shown in figure :9 by Tang *et al.* [4]

**Magneto refrigerant material**

This field of research is growing fast due to the characteristics of high conversion efficiency and ecological nature of magnetic cooling technology [247,266]. A magneto caloric study [152] in PNMO reveals relative cooling power of 41.35 $JKg^{-1}$ at 20 KOe and $\Delta S_{max}$ at 215K ($T_C$ = 213K) are 2.4 $JKg^{-1}K^{-1}$ and 4.9 $JKg^{-1}K^{-1}$ for 20 KOe and 50 KOe respectively. Zhang *et al.*[88] reported the reversible magnetic caloric effect around $T_C$ of ~208 K is correlated to second order magnetic transition. $\Delta S_{max}$ value of 1.25 $J/kg^{-1}K^{-1}$ and RCP of 48.2 J/kg for $\Delta H$ in the range 0–2T and $\Delta S_{max}$ value 2.48 $J/kg^{-1}K^{-1}$ and RCP of 150.1 J/kg for $\Delta H$ of 0–5 T. A study of PNMO and PCMO by changing the ratio of Co/Ni enabled the design of magnetocaloric composite material for use as magnetocaloric refrigerant in a wide temperature range [88].

**Spintronics**

Spintronics is thought to have huge potential for the next generation of information technology, since information can be transmitted at higher speeds and with low energy consumption [267]. Magnetodielectric property and the spin phonon coupling is pointing out that PNMO can be utilized for the spintronics applications [167]. Even though studies reveal that the spin phonon coupling is weak in PNMO, further advance research in the application domain is not extensive. The important behaviour that PNMO is having for exhibiting the magneto dielectric property is ferromagnetic insulator with positive superexchange interaction. The bulk properties exhibited to be reproduced in thin films in order to make the PNMO based spintronics device become a reality. For the application in magneto electronics as non-volatile memories studies on perovskites has been done in order to explore the coupled electric and magnetic properties. This magneto dielectric or magneto capacitance are used in spintronics applications. For novel devices like multiple-state memory elements, electric-field controlled magnetic sensors etc the simultaneous existence of ferromagnetism and ferroelectricity with spin phonon and spin polar coupling is necessary.

In the domain of gas sensing, biological application and the research on LNMO and some of the members of RNMO have been carried out and show attractive results as well. This gives an insight to explore other members of the RNMO family to these applications. For the adsorption of bovine serum albumin (BSA) application study has revealed that LNMO shows a good absorbing ability to BSA [268]. These are a potential carrier for large biomolecules, which will be widely used in the biomedical field and in biotechnology. Gas sensing applications in LNMO have been done with the help of impedance measurement under various gaseous environments with varied concentration. Magneto impedance at the IT magnetic field shows a positive value for oxygen, negative value for argon and nearly zero for nitrogen atmosphere [269]. The various industrial applications are possible with this compound including non-volatile memories, capacitors, sensors, actuators, resonant wave devices (such as radio- frequency filters), infra-red detectors, optical switches, and electric-motor overload protection circuits with multiple function.[270].

## Conclusion and future prospect

In this review paper various studies on PNMO that have been done in different parts of the world are reported. While there are lots of questions that still need to be answered. Novel predictions obtained from the theoretical calculations that call for experimental confirmation regarding rare earth double perovskites are

(i) The differences in response in terms of qualitative and quantitative aspects of various physical properties to hydrostatic pressure and chemical pressure.

(ii) How hydrostatic and chemical pressure depends on the antipolar displacements.

So, validation of these predictions through experimental analysis is needed. From Analysing the various studies on RNMO and specifically PNMO the rare earth ionic radius influences the Mn-0-Ni bond angle which correlates with the superexchange interaction. Hence Doping the A' site and/or B' site of PNMO can lead to changes in the magnetic, electric and spin phonon coupling properties. When an additional magnetic ion is doped either in A site and/or B'/B" site of the PNMO the changes in the magnetic behaviour and the variations in the conduction mechanism as well as the dielectric study after the doping are yet to be done with this material. The Neutron diffraction and soft x-ray scattering is warranted to study the temperature dependence of the structural properties and the ion positions, to explore the origin of the spin–phonon coupling obtained from the Raman study of PNMO. Neutron diffraction studies are necessary for the confirmation of the ordering of double perovskites, which strongly influence magnetic and electric properties. Studies on PNMO based on neutron diffraction are very few. If studies are extended in that direction as well, there will be many possibilities to unveil the vast and unique properties of the PNMO double perovskite. The low temperature spin glass behaviour of PNMO samples is not studied much in detail. The studies are done only as a comparison with other RNMO samples in this field of magnetism. Hence the glassy nature and the canted spins of the lattice and the coupling of Ni Mn an Pr magnetic ions leads to much complex magnetic behaviour. The study on these magnetic interactions provides insight to the richer phase diagram of these PNMO materials for both experimental and theoretical domain. The variable valence state of Ni and Mn leads to additional exchange interactions which will affect the phase transitions. So, the much-detailed study of PNMO with the help of XPS, EXAFS together with the magnetic study provide details about the local atomic structure and the influence of the valence state with the magnetic interactions in this particular member of RNMO family. Electrical properties with the determination of band gap and the conduction mechanisms need further studies in PNMO since these studies particularly on PNMO is very less. A site ordering of this family of double perovskites also needs attention for comparing the modulation of properties in these different arrangements. The thorough understanding of the spin as well as oxidation states of Ni and Mn cations is necessary to study the magnetic response of these systems. X-ray-absorption spectroscopy (XAS), which probes into the local atomic structure can be used to explore the short-range order and the exciting characteristics behind the PNMO double perovskite. These techniques which give details about the local structure of PNMO samples can also be utilized for further understanding of the exciting world of this particular double perovskite material. Hence detailed study of the structure correlation of PNMO and other members of the RNMO family is important in terms of probing the useful properties which call for practical applications.